\documentclass[a4paper,fleqn]{cas-dc}
\usepackage[authoryear]{natbib}
\graphicspath{{./}}

\def\tsc#1{\csdef{#1}{\textsc{\lowercase{#1}}\xspace}}
\tsc{WGM}
\tsc{QE}

\usepackage{amsmath}	
\usepackage{amssymb}	
\usepackage{tabularx}
\usepackage{graphicx}	
\graphicspath{ {figures/} {figures/template/} }
\usepackage{subfigure}	
\usepackage{hyperref}
\usepackage{float}
\usepackage{listings}
\usepackage{xcolor}
\usepackage{pythonhighlight}
\usepackage{ulem}
\usepackage[utf8]{inputenc}

\newcommand{\fargopy}{\href{https://fargopy.readthedocs.io}{\texttt{FARGOpy}}}
\newcommand{\plainfargopy}{\texttt{FARGOpy}}
\newcommand{\fargotd}{\texttt{FARGO3D}}
\newcommand{\webfargopy}{\href{https://fargopy.readthedocs.io}{\plainfargopy}}

\newcommand{\thetitle}{
Three-dimensional circumplanetary flows in a PDS~70c-inspired system: hydrodynamic simulations with \fargotd\ and analysis with \fargopy
}
\newcommand{\theshorttitle}{
High-Resolution 3D Simulations of the PDS 70c Circumplanetary Disk
}

\definecolor{mygreen}{RGB}{9,1,122}
\definecolor{myblack}{RGB}{80,30,20}

\ExplSyntaxOn
\keys_set:nn { stm / mktitle } { nologo }
\ExplSyntaxOff

\begin{document}
\let\WriteBookmarks\relax
\def\floatpagepagefraction{1}
\def\textpagefraction{.001}

\shorttitle{\theshorttitle}    

\shortauthors{Murillo-Gonzalez, A., Zuluaga, J.I. \& Montesinos, M.}  

\title[mode = title]{\thetitle}  

\author[1]{Alejandro Murillo-Gonzalez}[orcid=0009-0005-4433-2474]
\cormark[1]
\ead{bmurillo@usm.cl}
\credit{Design, preparation, and analysis of the simulations. Development of the interpolation and flux calculation modules of FARGOpy}

\author[2]{Jorge I. Zuluaga}[orcid=0000-0002-6140-3116]
\cormark[1]
\ead{jorge.zuluaga@udea.edu.co}
\credit{Identification of the scientific problem. Original conception, design and development of the FARGOpy package. Scientific interpretation of the results}

\author[1]{Matias Montesinos}[orcid=0000-0001-9789-5098]
\ead{matias.montesinosa@usm.cl}
\credit{Physics of the scientific problem. Review of initial conditions and solution of issues with the solution. Original conception, design and development of the FARGOpy package. Scientific interpretation of the results}

\affiliation[1]
            {organization={Departamento de F\'{\i}sica, Universidad T\'ecnica Federico Santa Mar\'ia},
            addressline={Avenida Espa\~na 1680}, 
            city={Valpara\'iso},
            citysep={}, 
            postcode={}, 
            country={Chile}}

\affiliation[2]
            {organization={SEAP/FACom, Instituto de F\'{\i}sica - FCEN, Universidad de Antioquia},
            addressline={Calle 70 No. 52-21}, 
            city={Medell\'in},
            citysep={}, 
            postcode={050026}, 
            country={Colombia}}

\cortext[1]{Corresponding author}

\begin{abstract}
The detection of circumplanetary material around the forming planet PDS~70c provides a valuable setting for studying three-dimensional gas accretion during giant-planet formation, while its vertical structure and mass transport remain poorly constrained. We present three-dimensional hydrodynamic simulations with \fargotd\ for a locally isothermal protoplanetary disk. The stellar mass, planetary orbital radius, disk aspect ratio, and explored planetary masses are motivated by the inferred properties of PDS~70c. Our aim is to determine how planetary mass controls the morphology, mass exchange, and rotational support of the circumplanetary gas under the same system-motivated disk conditions. We consider planets of $2$, $4$, and
$8,M_{\rm Jup}$, all above the local thermal mass for the adopted disk structure.

The locally isothermal prescription provides a controlled
prescribed-temperature limit that isolates the hydrodynamic organization of the flow, but does not describe the self-consistent thermal evolution or radiative observables of the circumplanetary gas. Gas enters the planetary environment through meridional flows from the upper disk layers together with streams near the midplane. Across the explored mass range, the flow progresses from a vertically extended, envelope-like configuration toward a more compact and stratified circumplanetary structure. The inferred CPD radius
increases from $R_{\rm CPD}\simeq0.31$ to $0.40,R_{\rm Hill}$, while the flow remains sub-Keplerian, reaching $v_\phi/V_{K,p}\sim0.79$--$0.85$. Most of the enclosed gas mass is concentrated within a few tenths of the Hill radius, whereas the outer Hill sphere is dominated by anisotropic inflow, outflow, and mass recirculation.

The three-dimensional analysis was performed with \fargopy, a Python-based post-processing framework developed and presented in this work to reconstruct three-dimensional fields, compute integrated fluxes, and characterize circumplanetary kinematics from \fargotd\ outputs. Thus, in addition to providing a hydrodynamic baseline for a PDS~70c-like circumplanetary environment, this study introduces \fargopy\ as a reproducible and reusable analysis framework for three-dimensional \fargotd\ simulations, while identifying which structural and kinematic properties can be constrained before including self-consistent radiative thermodynamics.

\end{abstract}



\maketitle

\section{Introduction}
\label{sec:introduction}

The formation of giant gaseous planets remains one of the central problems in planet formation theory. In the core-accretion scenario, once a protoplanetary core reaches a mass of order $\sim 10$--$20\,M_\oplus$, it enters a phase of rapid gas accretion from the surrounding protoplanetary disk (PPD) \citep{Bodenheimer-1986, Pollack-1996}. In this regime, the gas captured within the Hill sphere does not generally fall directly onto the planet but is redistributed into a circumplanetary structure that may take the form of a rotationally supported circumplanetary disk (CPD) or a pressure-supported envelope \citep{Szulagyi-2016}. The properties of this circumplanetary environment regulate the flow of mass and angular momentum onto the planet and provide the conditions under which regular satellites may form \citep{Canup-2002, Mosqueira-2003, Sasaki-2010, Szulagyi-2014, Szulagyi-2017}.

From the observational point of view, circumplanetary material is intrinsically difficult to isolate because of its compact size and its location inside structured transition disks. Nevertheless, recent advances in high-angular-resolution observations have made it possible to identify a small number of systems in which the circumplanetary environment can be studied indirectly through accretion tracers and compact dust emission. Among them, PDS~70 has become the benchmark case: it hosts two confirmed protoplanets embedded in the cavity of a transition disk, and PDS~70c, in particular, exhibits observational evidence of ongoing accretion and associated circumplanetary material \citep{keppler-2018, muller-2018, haffert-2019, wagner-2018, hashimoto-2020, isella-2019, benisty-2021}. This makes PDS~70c one of the best current laboratories for confronting hydrodynamic models of circumplanetary flows with observationally motivated conditions.

On the theoretical side, two-dimensional simulations have successfully described several large-scale properties of planet-disk interaction, including gap opening, cavity morphology, and the redistribution of gas and dust in transition disks \citep{Crida-2006, duffell-2015, Muley-2019, bae-2019}. Furthermore, the inclusion of radiative feedback in 2D models has been shown to significantly enhance the accretion rate and modify the thermal infrared signature of the circumplanetary region \citep{Montesinos+2015}. However, by construction, such models cannot capture the vertical structure of the gas, meridional circulation, or the three-dimensional flow topology within the planetary Hill sphere.

Three-dimensional simulations have shown that circumplanetary accretion is intrinsically non-planar, with gas often entering the Hill sphere through high-latitude flows and being redistributed through a combination of rotation, recirculation, and vertical motions \citep{Tanigawa-2012, Szulagyi-2014, Szulagyi-2016, Szulagyi-2017, fung-2019, li-2023}. Recent near-isothermal simulations by \citet{Sagynbayeva2025}, covering planetary masses up to $3\,M_{\rm Jup}$ and several protoplanetary-disk aspect ratios, further showed that the transition from envelope-like to disk-like circumplanetary structures occurs continuously and depends on both planetary mass and disk thickness. In parallel, radiation-hydrodynamic studies have demonstrated that the thermodynamic regime strongly affects whether the gas settles into a rotationally supported CPD or remains in a thicker, pressure-supported envelope \citep{Szulagyi-2016, krapp-2024}, a regime where radiative feedback further increases accretion rates, particularly within the ionization boundary \citep{Montesinos+2025}.

These studies establish the general three-dimensional nature of circumplanetary accretion and the dependence of CPD morphology on planetary and disk properties. The aim of the present work is therefore not to demonstrate these general trends again, but to characterize them quantitatively under a single global setup motivated by PDS~70c. We adopt the stellar mass and planetary orbital radius of the system, a disk aspect ratio consistent with models of the PDS~70 cavity, and a planetary-mass range of $2$--$8\,M_{\rm Jup}$. This range extends above that considered in recent near-isothermal parameter surveys and spans representative current estimates for the mass of PDS~70c. Within this common system-motivated framework, we examine how planetary mass modifies the vertical gas structure, the distribution of inflow and outflow across the Hill sphere, the enclosed circumplanetary mass, and the degree of rotational support.

In this work, we present high-resolution three-\\ dimensional hydrodynamical simulations of a PDS~70c-like system performed with \fargotd\ \citep{Benitez-Llambay2016}. The principal research question is how the morphology, mass exchange, and kinematics of the circumplanetary gas vary with planetary mass when the global circumstellar-disk conditions are kept fixed. We adopt a locally isothermal description of the gaseous disk and explore planetary masses of $2$, $4$, and $8\,M_{\rm Jup}$. We quantify the three-dimensional flow geometry, the mass flux through planet-centered control surfaces, the gas mass enclosed within different fractions of the Hill sphere, the regions of vertical supersonic motion and compression, and the planet-centered rotational kinematics. Rather than attempting to model the full thermodynamic complexity of the system, we focus on isolating the three-dimensional hydrodynamic organization of the flow in a controlled framework that is directly motivated by the observed architecture of PDS~70c. The simulations should therefore be interpreted as a hydrodynamic baseline for this system rather than as a complete radiative or thermochemical model.

A second contribution of this work is methodological. In order to analyze the three-dimensional outputs of \fargotd\ in a reproducible and physically meaningful way, we developed \fargopy, a Python-based post-processing framework that enables the reconstruction of three-dimensional fields, the extraction of arbitrary Cartesian slices, the computation of fluxes through closed surfaces, and the characterization of circumplanetary kinematics from simulation data. In this sense, the present study provides a quantitative characterization of the circumplanetary environment under PDS~70c-motivated conditions and introduces a tool designed to facilitate similar analyses in future studies and broaden the practical use of \fargotd\ within the community.

This paper is organized as follows. In \autoref{sec:pds70c} we summarize the observational context of the PDS~70 system. In \autoref{sec:numerical_method} we describe the hydrodynamical framework and the treatment of the planetary potential, while the simulation setup and initial conditions are presented in \autoref{sec:numerical_setup}. The post-processing methodology based on \fargopy\ is introduced in \autoref{sec:postprocessing}. The main results are presented in \autoref{sec:results}, discussed in
\autoref{sec:discussion}, and summarized in \autoref{sec:conclusions}.

\section{Observational constraints on PDS~70c}
\label{sec:pds70c}

PDS~70 is a K7 T~Tauri star with a stellar mass of $M_\star \simeq 0.76\,M_\odot$, an age of $\sim 5.4~\rm{Myr}$, and a distance of $\sim 113.43$~pc, surrounded by a pre-transitional disk with a large cavity \citep{dong-2012, Hashimoto+2012}. Its protoplanetary disk is one of the best-studied transition disks known, exhibiting a large inner cavity and a bright outer ring in the millimeter continuum \citep{long-2018, keppler-2019, benisty-2021}. Molecular-line observations further show that the cavity is not devoid of gas: the gaseous component extends well within the dust-depleted region, indicating that the planets orbit in a structured but not empty environment \citep{long-2018, keppler-2019, facchini-2021}. This combination of a cleared dust cavity, residual gas, and embedded planets makes PDS~70 a particularly valuable system for studying planet-disk interaction during the late stages of giant planet formation.

The system hosts two confirmed protoplanets, PDS~70b and PDS~70c, both embedded within the cavity \citep{keppler-2018, muller-2018, haffert-2019, christiaens-2019}. The outer companion, PDS~70c, lies at an orbital radius of about $34$~au and has become a key target for circumplanetary studies because it shows clear signatures of active accretion. In particular, hydrogen recombination-line emission, including H$\alpha$, has been detected at the location of the planet, indicating ongoing gas accretion onto its immediate environment \citep{haffert-2019}. The interpretation of these emission signatures requires careful consideration of the immediate planetary environment. In radiative-feedback models, an ionized envelope may affect the effective emitting region associated with accretion luminosity and $H_{\alpha}$ emission \citep{Montesinos+2025}. Current dynamical and observational constraints place its mass in the giant-planet regime, likely at a few Jupiter masses, although the exact value remains uncertain and model-dependent \citep{muller-2018, wang-2021, benisty-2021, portilla-revelo-2023}.

The strongest observational evidence for circumplanetary material in the system has been obtained around PDS~70c. ALMA observations at $855\,\mu$m revealed a compact continuum source spatially coincident with the planet and clearly separated from the main circumstellar dust ring \citep{isella-2019, benisty-2021}. This compact emission has been interpreted as dust associated with a circumplanetary disk, with an inferred characteristic size of order $\sim 0.3$--$0.5$~au and a dust mass that depends on the assumed temperature and opacity \citep{benisty-2021}. Three-dimensional radiative-transfer modeling further supports the presence of a compact and optically thick dusty CPD around PDS~70c, embedded in the inner edge of the outer disk ring \citep{portilla-revelo-2021}. Taken together, these observations strongly suggest that PDS~70c is surrounded by a circumplanetary structure capable of trapping solids while remaining connected to the gaseous cavity of the protoplanetary disk.

Recent high-angular-resolution observations have also provided evidence that the immediate environment of PDS~70c is dynamically structured. In particular,recent observations have revealed spiral-like or stream-like features linking the inner cavity with the vicinity of the planet, consistent with the idea that gas is being funneled toward the circumplanetary region through non-axisymmetric flows \citep{christiaens-2024}. Similar flow morphologies have also been predicted in numerical models of the PDS~70 system \citep{toci-2020}. Although these observations do not directly resolve the full three-dimensional gas dynamics within the Hill sphere, they strongly suggest that the environment of PDS~70c is shaped by a combination of accretion streams, circumstellar-disk interaction, and compact circumplanetary material.

For the purposes of this work, the most relevant observational constraints are therefore the stellar mass of the system, the orbital location of PDS~70c within a gas-bearing transition-disk cavity, the evidence for ongoing accretion onto the planet, and the presence of compact circumplanetary material inferred from millimeter observations and radiative-transfer modeling. These elements provide the empirical basis for the system-motivated hydrodynamical setup adopted below, where we focus on the three-dimensional gas flow in the circumplanetary environment of PDS~70c under controlled physical assumptions.

\section{Hydrodynamic simulations in \fargotd}
\label{sec:numerical_method}

To model the three-dimensional gas dynamics in the circumplanetary environment, we use \fargotd\ \citep{Benitez-Llambay2016}, an open-source hydrodynamical code designed to solve the equations of compressible gas dynamics in rotating disk systems. In this work, the fluid is treated as viscous and non-self-gravitating, and the equations are solved in a non-inertial frame co-rotating with the planet.

In conservative form, the hydrodynamic equations solved by \fargotd\ can be written as follows:
\begin{eqnarray}
\partial_t \rho
+ \nabla \cdot (\rho \mathbf{v}) & = & 0\\
\partial_t (\rho \mathbf{v}) + \nabla \cdot \left(\rho \mathbf{v} \otimes \mathbf{v}\right) + & & \\
+ \nabla \cdot \left(P \mathbf{I} - \mathbf{T} \right)
& =
& - \rho \nabla \Phi_{\rm tot} + \nonumber\\
& & - 2 \rho\, \boldsymbol{\Omega}_{\rm f} \times \mathbf{v}  + \nonumber\\
&& - \rho\, \boldsymbol{\Omega}_{\rm f} \times
\left( \boldsymbol{\Omega}_{\rm f} \times \mathbf{r} \right),
\label{eq:momentum}
\end{eqnarray}
where $\otimes$ is the dyadic product, $\rho$ is the gas volume density, $\mathbf{v}$ is the velocity field, $P$ is the pressure, $\mathbf{T}$ is the viscous stress tensor, $\Phi_{\rm tot}$ is the total gravitational potential, and $\boldsymbol{\Omega}_{\rm f}$ is the angular velocity of the rotating frame. The viscous stress tensor is written using an $\alpha$-viscosity prescription \citep{shakura-1973}.

We adopt a locally isothermal thermodynamic closure in which the thermal structure of the disk is prescribed, and no energy equation is evolved. The pressure is therefore written as
\begin{equation}
P = c_{\rm s}^{2}(R)\,\rho,
\label{eq:isothermal_eos}
\end{equation}
where $c_s(R)$ is the prescribed sound speed. Here $R=r\sin\theta$ is the cylindrical radius, whereas $r$ denotes the spherical radius measured from the star. This approximation does not aim to capture the full thermodynamic evolution of the circumplanetary gas but instead provides a controlled framework in which the three-dimensional hydrodynamic organization of the flow can be isolated and analyzed.

We solve the hydrodynamic equations with \fargotd\
\citep{Benitez-Llambay2016}. For the azimuthal transport, we employ
\textsc{RAM} (Rapid Advection Algorithm on Arbitrary Meshes;
\citealt{benitez-llambay-2023}), which extends the orbital-advection approach
to non-uniform meshes and improves the efficiency and accuracy of rapidly
rotating flows.

\subsection{Planetary potential and smoothing}
\label{subsec:planet_potential}

The planet is modeled as a point mass orbiting within the disk. In the immediate vicinity of the planet, however, the point-mass potential becomes singular, producing artificially large accelerations and severe timestep limitations. Moreover, the physical scales associated with the planetary radius and its innermost gaseous environment are not explicitly resolved on the hydrodynamical grid. For these reasons, the planetary potential is regularized through a Plummer-like softening prescription:
\begin{equation}
\Phi_p(\mathbf{r}) = - \frac{G M_p}{\sqrt{|\mathbf{r}-\mathbf{r}_p|^2 + \epsilon^2}},
\label{eq:plummer_potential}
\end{equation}
where $M_p$ is the planetary mass, $\mathbf{r}_p$ is the position of the planet, and $\epsilon$ is the smoothing length. We adopt this potential through the
\texttt{RocheSmoothing} prescription in \fargotd. This regularization avoids
the singularity at the planetary position and permits a smooth reduction of
the smoothing length during the initial relaxation. We note that its influence
extends beyond $\epsilon$, and therefore the innermost flow should be
interpreted in the context of the adopted softened potential.

In the \texttt{RocheSmoothing} scheme of \fargotd, the smoothing length is expressed as a fraction of the Hill radius,
\begin{equation}
\epsilon = S\,R_{\rm Hill}, \qquad
R_{\rm Hill} = a_p \left( \frac{M_p}{3M_\star} \right)^{1/3},
\label{eq:hill_radius}
\end{equation}
where $a_p$ is the orbital radius and $S$ is a dimensionless smoothing parameter. This formulation links the regularization scale to the characteristic size of the planet's gravitational sphere of influence.

Because the abrupt introduction of a planet with its full mass and final smoothing scale can generate strong numerical transients, we adopt a time-dependent smoothing strategy. The parameter $S$ evolves smoothly from an initial value $S_{\rm ini}$ to a final value $S_{\rm fin}$ over a transition interval of $N_{\rm tr}$ planetary orbits:
{\small
\begin{equation}
S(N_{\rm orb}) =
\left\{
\begin{array}{ll}
S_{\rm ini}, & N_{\rm orb} \le 0, \\[4pt]
S_{\rm fin} + (S_{\rm ini}-S_{\rm fin})\,W(N_{\rm orb}),
& 0 < N_{\rm orb} < N_{\rm tr}, \\[4pt]
S_{\rm fin}, & N_{\rm orb} \ge N_{\rm tr}.
\end{array}
\right.
\label{eq:smoothing_transition}
\end{equation}
}
where $N_{\rm orb}$ is the number of elapsed planetary orbits and $W(N_{\rm orb})$ is a smooth transition function. Following \citet{Szulagyi-2016}, we adopt
\begin{equation}
W(N_{\rm orb}) = \cos^4\!\left(\frac{\pi}{2}\,\frac{N_{\rm orb}}{N_{\rm tr}}\right),
\label{eq:transition_function}
\end{equation}
which guarantees a continuous evolution of the softened potential and minimizes spurious forcing of the flow during the early stages of the calculation.

We complement this treatment with the \texttt{masstaper} option of \fargotd, which gradually increases the planetary mass from zero to its target value:
\begin{equation}
M_p(t) = M_{p,\mathrm{final}}\,f(t),
\label{eq:masstaper}
\end{equation}
where $f(t)$ is a smooth growth function. The purpose of this procedure is not to model the physical growth history of the planet but to allow the circumplanetary flow to adjust progressively to the planetary potential and to reduce artificial transients associated with the initialization.

The combination of a softened planetary potential, a time-dependent smoothing prescription, and a gradual mass taper provides a numerically robust framework for studying the three-dimensional structure of circumplanetary gas. The specific parameter values adopted in the simulations presented in this work are given in \autoref{sec:numerical_setup}.

Because the physical radius of the planet is unresolved, the
simulations do not follow the final incorporation of gas onto the planetary
surface. At late times, the gas mass enclosed within $R_{\rm Hill}$ amounts
to only $\simeq0.22$--$0.32\%$ of $M_p$ across the three models. The
resolved circumplanetary gas therefore remains dynamically negligible
compared with the planetary mass and does not significantly affect the
large-scale flow dynamics.

\section{Numerical setup}
\label{sec:numerical_setup}

The simulations are designed to represent a PDS~70c-like system under controlled physical assumptions. The adopted setup is guided by the stellar mass of PDS~70, the orbital location of PDS~70c within the disk cavity, and previous observationally motivated estimates of the disk aspect ratio and large-scale gas distribution. The numerical parameters used in the simulations are summarized in \autoref{tab:numerical_setup}.

\begin{table*}
\centering
\begin{tabular}{l c l}
\hline\hline
\multicolumn{3}{c}{Observed or system-motivated quantities} \\
\hline
Stellar mass & $M_\star$ & $0.76\,M_\odot$\textsuperscript{a} \\
System age &  & $\sim 5.4\pm 1.0$ Myr\textsuperscript{a} \\
Distance &  & $113.43$ pc\textsuperscript{a} \\
PDS~70c orbital radius & $r_p$ & $34$ au\textsuperscript{a} \\
Considered planetary masses & $M_p$ & $2,\ 4,\ 8\,M_{\rm Jup}$ \\
\hline
\multicolumn{3}{c}{Adopted disk model} \\
\hline
Surface density profile & $\Sigma(R)$ & $\Sigma_0\,R^{-1}$\textsuperscript{b} \\
Flaring index & $f$ & $0.30$\textsuperscript{b} \\
Aspect ratio at $r_p$ & $H/R$ & $0.072$\textsuperscript{b} \\
Turbulent viscosity & $\alpha$ & $5\times10^{-3}$ \\
Equation of state &  & Locally isothermal \\
\hline
\multicolumn{3}{c}{Numerical setup} \\
\hline
Coordinate system &  & Spherical $(r,\theta,\phi)$ \\
Radial domain &  & $0.4 \le r \le 2.0\ (13.6$--$68$ au) \\
Azimuthal extent &  & $-\pi \le \phi < \pi$ \\
Vertical domain &  & Half-disk ($\theta_{\rm min}\le\theta\le\pi/2$) \\
Reference frame &  & Co-rotating with the planet \\
Code units &  & $G=1,\ M_\star=1,\ r_p=1$ \\
Initial smoothing & $S_{\rm ini}$ & $0.30$ \\
Final smoothing & $S_{\rm fin}$ & $0.12$ \\
Smoothing transition & $N_{\rm tr}$ & $100$ orbits \\
Planetary mass growth &  & \texttt{masstaper}, 10 orbits \\
Medium resolution & $(N_\phi,N_r,N_\theta)$ & $(352,224,64)$ \\
High resolution & $(N_\phi,N_r,N_\theta)$ & $(544,320,96)$ \\
Radial cells within $2\,R_{\rm H}$ (MR/HR) &  & $91\ /\ 129$ \\
Azimuthal cells within $2\,R_{\rm H}$ (MR/HR) &  & $82\ /\ 126$ \\
Total cells within $V=2/3\pi R_{\rm H}^3$ (MR/HR) &  & $5814\ /\ 12826$ \\
Total integration time &  & $200\ \mathrm{PDS}~70c\ \mathrm{orbits}\ (4.55\times10^{4}~\rm{yr})$ \\
\hline\hline
\end{tabular}
\caption{Observed, adopted, and numerical parameters used in the three-dimensional hydrodynamical simulations of a PDS~70c-like system.}
\label{tab:numerical_setup}
\vspace{2mm}
{\footnotesize
Notes.
\textsuperscript{a} Stellar properties, distance, and orbital parameters of the PDS~70 system from \citet{keppler-2018,muller-2018, keppler-2019}.
\textsuperscript{b} Adopted disk parameters motivated by hydrodynamical and thermochemical modeling of PDS~70 \citep{Muley-2019, portilla-revelo-2023}.
}
\end{table*}

\subsection{Computational domain and grid}
\label{subsec:computational_domain}

The simulations are performed in spherical polar coordinates $(r,\theta,\phi)$, modeling only the upper hemisphere of the disk under the assumption of reflection symmetry with respect to the midplane. The computational domain extends over $0.4 \le r \le 2.0$ in code units, spans the full azimuthal range $-\pi \le \phi < \pi$, and covers the colatitude interval $1.4~\mathrm{rad} \le \theta \le \pi/2$.

\begin{figure}
    \centering
    \includegraphics[width=\linewidth]{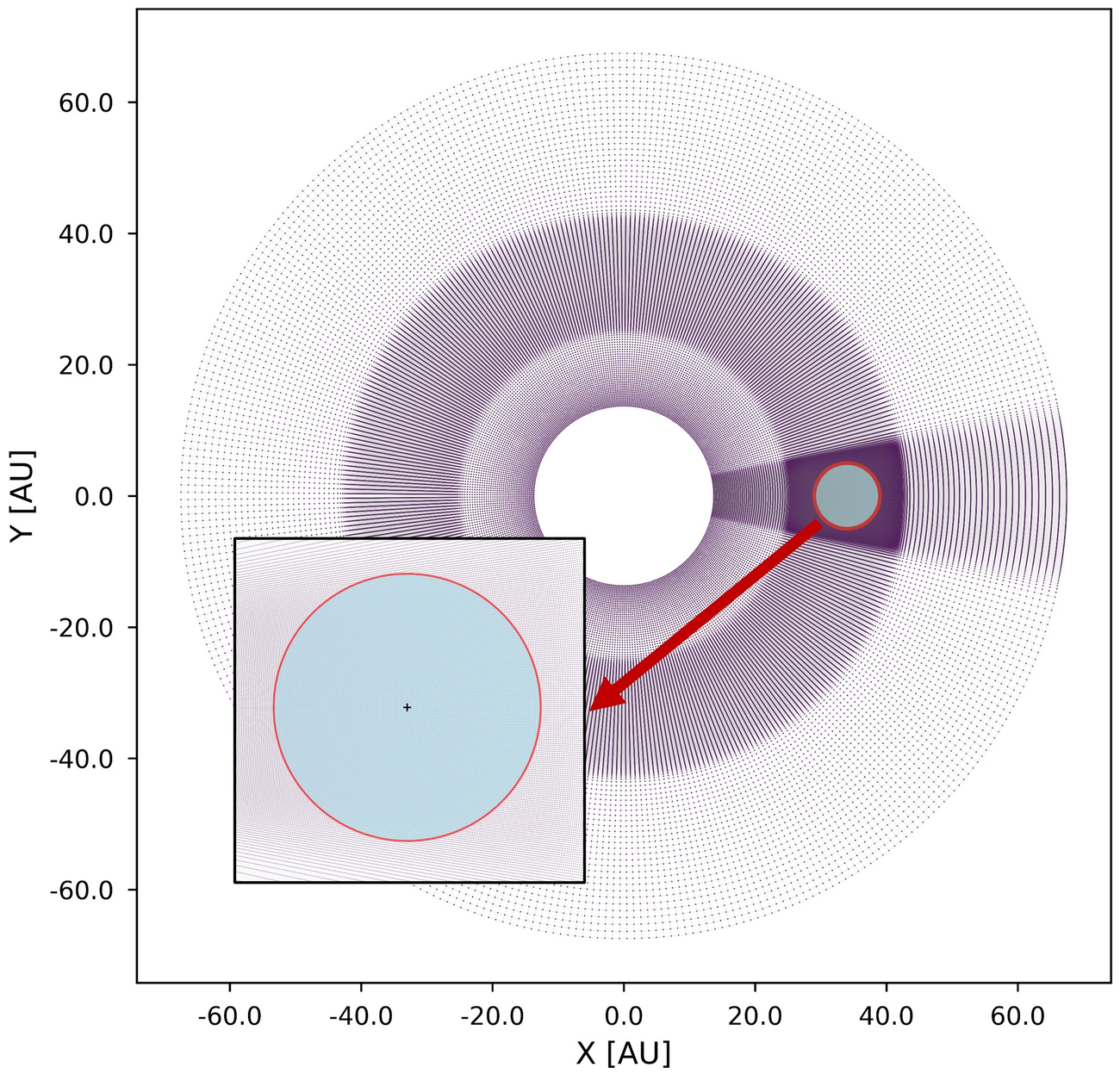}
    \caption{
    Distribution of the computational mesh in the orbital plane $(X,Y)$, showing the non-uniform refinement scheme used in the simulations. The point density illustrates the concentration of cells in the vicinity of the planetary orbit, while the red circle indicates the projection of the planet's Hill radius onto the midplane. This refinement ensures adequate sampling of the circumplanetary region, while maintaining a coarser resolution in the rest of the disk.
    }
    \label{fig:mesh}
\end{figure}

The mesh is globally non-uniform in the radial and azimuthal directions and is constructed following the mesh-density formalism of \citet{benitez-llambay-2023}. In practice, analytic mesh-density functions and their cumulative coordinate transformations are used to concentrate resolution smoothly around the orbital radius of the planet. A schematic illustration of the resulting grid is shown in \autoref{fig:mesh}. The parameters \texttt{XMa}, \texttt{XMb}, and \texttt{XMc} control the radial concentration of cells around $r=1$, whereas \texttt{YMa}, \texttt{YMb}, and \texttt{YMc} regulate the azimuthal concentration in the vicinity of the planet.

Two families of simulations were carried out, one at medium resolution and another at high resolution. They employ grids of $352 \times 224 \times 64$ and $544 \times 320 \times 96$ cells in $(N_\phi,N_r,N_\theta)$, respectively. In the orbital midplane, the Hill-sphere diameter is resolved by up to $91$ radial and $82$ azimuthal cells in the medium-resolution runs, and by $129$ radial and $126$ azimuthal cells in the high-resolution runs, corresponding to effective resolutions of $\sim 45$--$50$ and $\sim 65$ cells per $R_{\rm H}$, respectively. In volumetric terms, the number of cells within a sphere of radius $R_{\rm H}$ increases from $5814$ to $12826$. This refinement is sufficient to resolve the large-scale structure of the circumplanetary flow while maintaining its connection to the global circumstellar disk.

\subsection{Initial conditions}
\label{subsec:initial_conditions}

The simulated system is centered on a star of mass $M_\star = 0.76\,M_\odot$, and the planet is placed at an orbital radius of $r_p = 34$ au, consistent with the observed properties of PDS~70 and PDS~70c \citep{keppler-2018, muller-2018, haffert-2019, wang-2021}. The code uses dimensionless units such that $G = 1$, $M_\star = 1$, and $r_p = 1$, with time measured in units of $\Omega_p^{-1}$, where $\Omega_p = (G M_\star / r_p^3)^{1/2} = 1$.

The initial vertical structure of the protoplanetary disk is prescribed through the relative scale height $h(R)\equiv H/R$. Following \citet{Muley-2019}, we adopt
\begin{equation}
h(R) = h_0 \left(\frac{R}{R_0}\right)^f,
\end{equation}
with $h_0 = 0.063$ at $R_0 = 22$ au and $f = 0.30$. Extrapolated to the orbital radius of PDS~70c, this yields $(H/r)_{r_p} \simeq 0.072$. Under the locally isothermal approximation, the pressure is written as:
\begin{equation}
P(R,\theta) = \rho(R,\theta)\,c_{\rm s}^2(R),
\end{equation}
with sound speed
\begin{equation}
c_{\rm s}(R) = h(R)\,v_{\rm K}(R), \qquad
v_{\rm K}(R)=R^{-1/2}.
\end{equation}

The initial gas surface density is taken to follow
\begin{equation}
\Sigma(R)=\Sigma_0\,R^{-1},
\label{eq:sigma_profile_ic}
\end{equation}
and $\Sigma_0$ is fixed so that the adopted profile reproduces a representative minimum gas mass scale consistent with previous thermo-chemical estimates for the PDS~70 disk \citep{portilla-revelo-2023}. This gives $\Sigma_0 \simeq 1.6\times 10^{-5}$ in code units, equivalent to $\Sigma_0 \simeq 9.3\times10^{-2}\ \mathrm{g\,cm^{-2}}$ at 34 au. Assuming vertical hydrostatic equilibrium, the initial volumetric density profile is
\begin{equation}
\rho(R,z)=
\frac{\Sigma(R)}{\sqrt{2\pi}\,H(R)}
\exp\left[-\frac{z^2}{2H^2(R)}\right],
\label{eq:rho3d}
\end{equation}
where $z$ is measured from the midplane.

The kinematic viscosity is prescribed through the standard $\alpha$-prescription,
\begin{equation}
\nu(R)=\alpha\,c_{\rm s}(R)\,H(R),
\end{equation}
with a uniform value $\alpha = 5\times10^{-3}$, consistent with low-viscosity models proposed for the PDS~70 disk \citep{portilla-revelo-2023}. The three planetary masses explored in this work, $2$, $4$, and $8\,M_{\rm Jup}$, are chosen to span the range typically inferred for PDS~70c while allowing us to investigate how the circumplanetary flow changes across a representative giant-planet mass interval.

\subsection{Boundary conditions}
\label{subsec:boundaries}

At the radial boundaries, pressure-supported near-Ke-plerian extrapolation conditions are imposed for the density and azimuthal velocity. The ghost cells are filled, assuming the initial rotational equilibrium profile is corrected for radial pressure support. In particular, the azimuthal velocity is extrapolated using
\begin{equation}
v_\phi(R,\theta)=
v_{\rm K}(R)\sqrt{1-(p+2f)\,h^2(R)}
-\Omega_{\rm f} R \sin\theta,
\end{equation}
where $p$ is the radial surface-density exponent, defined through $\Sigma \propto R^{-p}$, and $f$ is the flaring index, defined through $h \propto R^f$. This treatment helps maintain the outer disk close to rotational equilibrium and reduces spurious reflections of spiral perturbations at the radial edges of the domain.

At the midplane, symmetry with respect to $\theta=\pi/2$ is imposed. The density and the velocity components $v_r$ and $v_\phi$ are taken to be symmetric, while the meridional component $v_\theta$ is antisymmetric, consistent with reflection symmetry across the disk midplane.

At the upper boundary, $\theta=\theta_{\rm min}$, the density is extrapolated assuming vertical hydrostatic equilibrium according to \autoref{eq:rho3d}, while the velocity components are treated with outflow conditions. This choice allows perturbations generated within the disk to leave the computational domain without introducing significant artificial reflections.

\section{Post-processing with \webfargopy}
\label{sec:postprocessing}

\subsection{Introducing \webfargopy}
\label{subsec:intro_fargopy}

The analysis of the three-dimensional hydrodynamical simulations carried out with \fargotd\ was performed using a fully reproducible post-processing workflow implemented in the Python package \webfargopy\footnote{For a comprehensive description of the package, including usage examples, API documentation, and an image gallery, see \href{https://fargopy.readthedocs.io}{https://fargopy.readthedocs.io}.} (see Appendix~\ref{app:fargopy}). The package was developed to analyze outputs defined on the native spherical mesh of \fargotd\ in a form suitable for circumplanetary diagnostics, including planar cuts, spherical control surfaces, enclosed-mass estimates, and planet-centered kinematic measurements.

In this work, \webfargopy\ is used as an evaluation and analysis layer on top of the simulation outputs. It enables coordinate transformations, interpolation of scalar and vector fields onto arbitrary grids or surfaces, and direct integration of diagnostic quantities relevant to the circumplanetary flow. Details of the interpolation scheme and validation tests are deferred to Appendix~\ref{app:fargopy_interpolation}.

\subsection{Specific postprocessing setup}
\label{subsec:fargopy_postprocessing}

Although the simulations were performed in spherical coordinates $(r,\theta,\phi)$, all diagnostics were evaluated in Cartesian coordinates $(X,Y,Z)$ obtained through explicit coordinate transformations of the hydrodynamical fields. Here, $(X,Y)$ denotes the orbital plane and $Z$ the vertical direction. This choice facilitates the construction of physically meaningful planar cuts and the analysis of the flow in a planet-centered frame.

\begin{figure*}
    \centering
    \includegraphics[width=\linewidth]{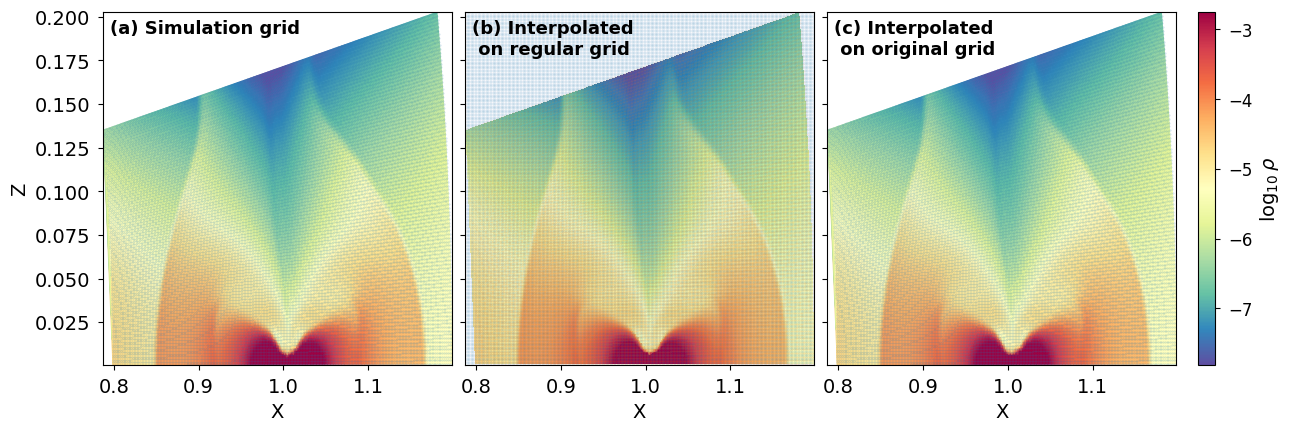}
    \caption{
    Visual comparison of the density field in a meridional slice.
    \textit{Left:} original field on the spherical simulation mesh.
    \textit{Center:} field interpolated onto a regular Cartesian grid using \fargopy.
    \textit{Right:} field reconstructed on the original mesh after the round-trip procedure.
    All panels use the same $\log_{10}\rho$ normalization.
    }
    \label{fig:roundtrip_density}
\end{figure*}

Field values were evaluated using \fargopy\ on target Cartesian grids using linear interpolation from the native spherical mesh. Interpolation in \fargopy\ is used strictly as an evaluation operator: it does not alter the dynamical state of the simulation but enables the sampling of the solution on arbitrary slices and surfaces required by the analysis. To assess interpolation fidelity, a round-trip reconstruction test was performed by interpolating fields from the spherical grid onto a Cartesian grid and then back onto the original mesh. The example shown in \autoref{fig:roundtrip_density} indicates that the adopted procedure in our package preserves the global morphology of the density field and does not introduce significant smoothing artifacts at the working resolution.

Two additional diagnostics were constructed in the meridional planes. The first is the vertical Mach number,
\begin{equation}
\mathcal{M}_Z = \frac{|v_Z|}{c_s},
\label{eq:mach_z}
\end{equation}
where $v_Z$ is the vertical Cartesian velocity component and $c_s(R)=h(R)\,v_K(R)$ is the prescribed local sound speed. Regions with $\mathcal{M}_Z \gtrsim 1$ identify vertical motions that exceed local thermal support and therefore mark potential sites of strong deceleration or compression. The second is a compression tracer based on the velocity divergence.
\begin{equation}
\mathcal{C} = \max(0,-\nabla\cdot\mathbf{v}),
\label{eq:compression_tracer}
\end{equation}
which isolates regions of convergent flow. Within a given slice, the divergence is evaluated from the in-plane velocity components, providing a practical diagnostic of compressive layers and shock-like structures in the circumplanetary environment. For visualization purposes only, a mild Gaussian smoothing was applied to maps of $\mathcal{C}$ in order to reduce pixel-scale granularity; all quantitative analyzses were performed on the unsmoothed fields.

\subsection{Mass flow}
\label{subsec:fargopy:mass_flows}

To quantify the exchange of mass between the circumstellar disk and the circumplanetary region, we evaluated mass fluxes through spherical control surfaces centered on the planet. In an Eulerian framework, the rate of change of mass within a volume $V$ is related to the flux through its boundary $\partial V$ by
\begin{equation}
\frac{d}{dt}\int_V \rho\,dV = -\oint_{\partial V}\rho\,\mathbf{v}\cdot d\mathbf{S},
\label{eq:continuity_integral}
\end{equation}
where $d\mathbf{S}=\hat{\mathbf{n}}\,dS$ is the oriented surface element. Following this definition, \webfargopy\ constructs spherical surfaces of radius $R$, expressed as a fraction of the Hill radius $R_{\rm H}$, and discretizes them through a nearly uniform triangular tessellation (see Appendix~\ref{app:fargopy_tessellation}).

For each triangular element $i$, we compute its area $A_i$, its outward unit normal vector $\hat{\mathbf{n}}_i$, and the coordinates of its centroid. The density and velocity fields are then interpolated at that location, and the mass flux through the element is evaluated as
\begin{equation}
\dot{m}_i = \rho_i\,(\mathbf{v}_i\cdot\hat{\mathbf{n}}_i)\,A_i,
\label{eq:flux_element}
\end{equation}
such that $\dot{m}_i<0$ denotes inflow and $\dot{m}_i>0$ denotes outflow. The net mass flow rate through the spherical surface is obtained by summing over all elements.
\begin{equation}
\dot{M}(R) \approx \sum_i \dot{m}_i.
\label{eq:net_flux}
\end{equation}
This construction allows us to quantify both the magnitude and the angular distribution of the mass exchange across the Hill sphere.

In addition to surface fluxes, we estimated the gas mass enclosed in regions centered on the planet. Unlike the flux calculation, this quantity was obtained directly from the native \fargotd\ mesh rather than from interpolated fields. The enclosed mass in a volume $V$ is computed as
\begin{equation}
M(V) = \int_V \rho\,dV \simeq \sum_{j\in V}\rho_j\,\Delta V_j,
\label{eq:enclosed_mass}
\end{equation}
where $\rho_j$ is the density in cell $j$ and $\Delta V_j$ is the corresponding cell volume. In spherical coordinates,
\begin{equation}
\Delta V = r^2\sin\theta\,\Delta r\,\Delta\theta\,\Delta\phi.
\label{eq:volume_element}
\end{equation}
The selection of cells is performed in the planet-centered frame through geometric masks. For a spherical control volume of radius $R$, the mask is defined by
\begin{equation}
(x-x_p)^2+(y-y_p)^2+(z-z_p)^2 \le R^2.
\label{eq:spherical_mask}
\end{equation}
This procedure yields time series $M(R,t)$ that can be compared directly with the fluxes $\dot{M}(R)$ and with the local kinematics. A limitation of this diagnostic is that a control sphere does not distinguish between gravitationally bound gas and gas in transit; accordingly, the enclosed mass is interpreted together with the measured fluxes and the rotational properties of the flow.

\subsection{Kinematics}
\label{subsec:cpd_kinematics_map}

The rotational properties of the circumplanetary gas were quantified by
projecting the velocity field onto the planetocentric azimuthal direction in
the $XY$ plane. Given the relative position vector
$\mathbf{r}=(x-x_p,\,y-y_p)$ and the planetocentric azimuthal angle
$\phi=\tan^{-1}[(y-y_p)/(x-x_p)]$, the corresponding azimuthal unit
vector is
\begin{equation}
\hat{\boldsymbol{e}}_\phi
=
-\sin\phi\,\hat{\boldsymbol{x}}
+
\cos\phi\,\hat{\boldsymbol{y}},
\end{equation}
and the planetocentric azimuthal velocity is defined as
\begin{equation}
v_\phi
=
\mathbf{v}\cdot\hat{\boldsymbol{e}}_\phi.
\label{eq:vphi_planetocentric}
\end{equation}

To quantify the degree of rotational support, the azimuthal velocity
is normalized by the circular velocity associated with the softened planetary
potential adopted in the simulations. For the Plummer potential defined in
\autoref{eq:plummer_potential}, this velocity is
\begin{equation}
V_{K,p}(r)
=
\left(
r\frac{{\rm d}\Phi_p}{{\rm d}r}
\right)^{1/2}
=
\left[
\frac{G M_p r^2}
{\left(r^2+\epsilon^2\right)^{3/2}}
\right]^{1/2},
\label{eq:vk_planet}
\end{equation}
where $r=|\mathbf{r}|$ is the planetocentric distance in the orbital plane.
For the maps and radial profiles evaluated at 200 orbits, the
smoothing length is fixed at its final value,
$\epsilon=S_{\rm fin}R_{\rm Hill}=0.12\,R_{\rm Hill}$. At
$r\gg\epsilon$, \autoref{eq:vk_planet} recovers the point-mass Keplerian
velocity.

From this quantity, we constructed two-dimensional maps of
$v_\phi/V_{K,p}$ in the orbital plane and azimuthally averaged radial profiles,
\begin{equation}
\left\langle
\frac{v_\phi}{V_{K,p}}
\right\rangle(r),
\label{eq:vphi_profile}
\end{equation}
together with the 16--84 percentile interval\footnote{The 16--84 percentile
interval is used as an analog of the $\pm1\sigma$ range, without assuming a
Gaussian distribution of the velocity field, which may exhibit asymmetries
and tails associated with non-circular flows.} to quantify the azimuthal
dispersion. In this framework, values close to unity indicate
azimuthal velocities comparable to the circular velocity associated with the
softened planetary gravity. Systematic deviations reflect the combined effects
of pressure support, meridional motions, non-axisymmetric flows, stellar tidal
forcing, and angular-momentum transport.

Taken together, the post-processing strategy combines interpolation-based
visualization in physically meaningful planes, Eulerian measurements of fluxes
across spherical surfaces, direct integration of enclosed mass on the native
mesh, and planet-centered kinematic diagnostics. This combination allows the
circumplanetary gas to be characterized not only as a local overdensity but
also as a three-dimensional, dynamically connected flow structure.

\section{Results}
\label{sec:results}

To assess how planetary mass controls the formation and structure of the circumplanetary disk, we analyze the gas morphology, mass transport, and kinematics inside the Hill sphere. We first examine the flow in the orbital plane, where the degree of confinement and rotational organization can be identified most directly. We then characterize the vertical structure of the gas, the three-dimensional geometry of inflow and outflow, the time evolution of the mass budget, and the degree of centrifugal support in the circumplanetary region.

\subsection{Global morphology of the CPD}
\label{subsec:cpd_morphology}

\begin{figure*}
\centering
\includegraphics[width=\textwidth]{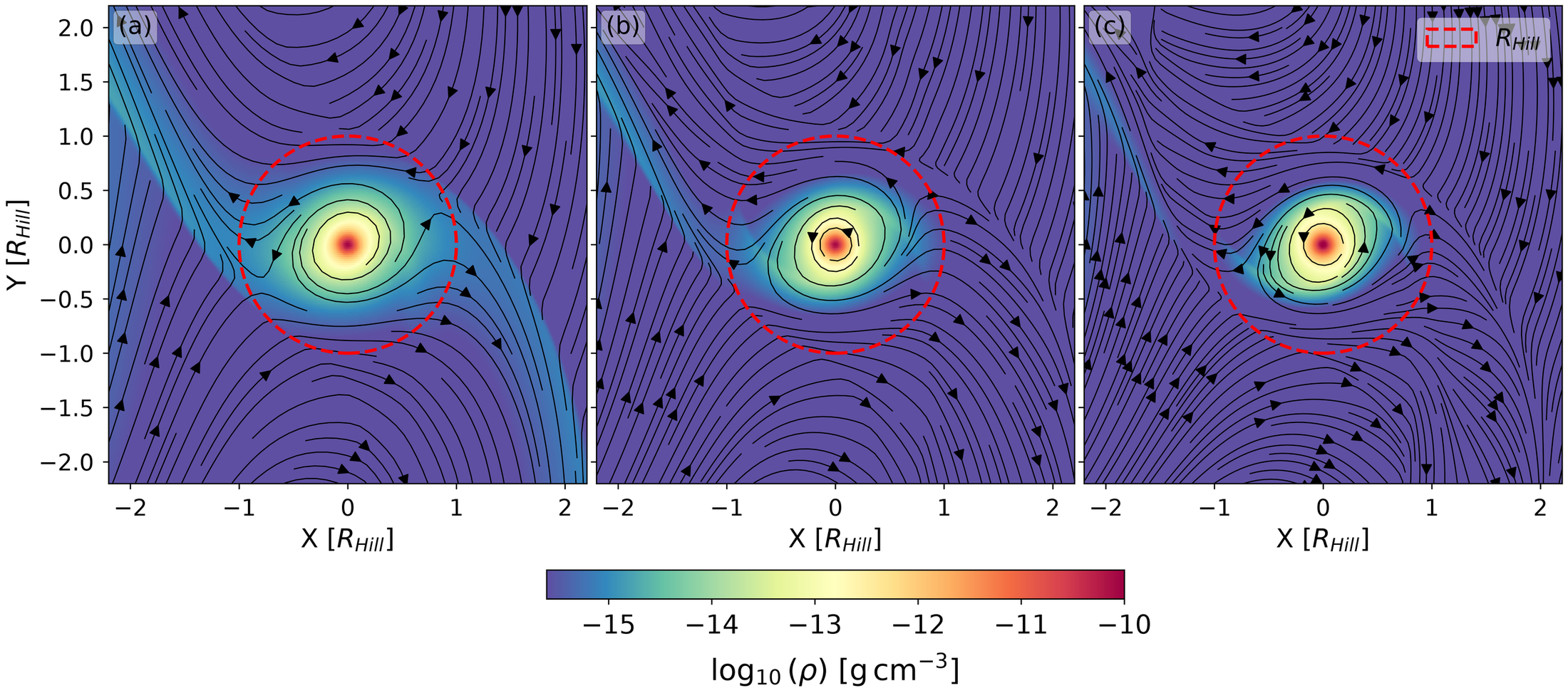}
\caption{Gas volume density in the orbital ($XY$) plane for planetary masses
of (a) $2\,M_{\rm Jup}$, (b) $4\,M_{\rm Jup}$, and (c) $8\,M_{\rm Jup}$.
The background shows $\log_{10}(\rho)$, and the black curves correspond to
instantaneous streamlines of the velocity field projected onto the
orbital plane. Distances are expressed in units of the Hill radius
$R_{\rm Hill}$. The star lies in the direction of negative $X$.}
\label{fig:xy_streamlines}
\end{figure*}

\autoref{fig:xy_streamlines} shows the midplane density structure and
the instantaneous velocity field projected onto the orbital plane
for the three planetary masses considered. The streamlines therefore
represent patterns of the projected Eulerian velocity field and should not be
interpreted as closed three-dimensional trajectories of individual gas
parcels.

The figure isolates the circumplanetary region within approximately one Hill
radius and reveals a clear progression from a weakly confined configuration
at low mass to a more compact and dynamically coherent structure at high mass.

For $M_{\rm p}=2\,M_{\rm Jup}$, the density enhancement inside
$R_{\rm Hill}$ is relatively extended and merges smoothly with the surrounding
circumstellar flow. Most streamlines in the projected midplane velocity
field remain open and reconnect with the background shear, while only a small
central region exhibits closed circulation patterns in projection.
In this case, the planetary potential perturbs the local flow but does not
establish a well-confined circumplanetary structure throughout most of the Hill
sphere.

For $M_{\rm p}=4\,M_{\rm Jup}$, the density peak becomes more centrally
concentrated, and the projected flow pattern changes qualitatively.
Closed circulation patterns occupy a larger fraction of the inner Hill
sphere, while gas approaching from the circumstellar disk is more efficiently
deflected around the circumplanetary region. This marks the transition toward
a more centrally concentrated and rotationally structured
configuration.

For $M_{\rm p}=8\,M_{\rm Jup}$, the circumplanetary structure becomes markedly
compact and centrally peaked. The projected inner velocity field
exhibits tightly wound circulation patterns, whereas the external flow is
strongly diverted around the dense core. In this regime, the circumplanetary
flow is more strongly confined and less affected by the local shear of
the circumstellar disk.

Taken together, the midplane morphology shows that increasing planetary mass
enhances the confinement of gas inside the Hill sphere, reduces the penetration
of circumstellar shear into the inner region, and increases the radial
extent of the circulation patterns identified in the projected velocity
field.

The increasing planetary mass also produces progressively deeper
gaps in the circumstellar disk. At $t=200$ orbits, the azimuthally averaged
surface density at the planetary orbital radius decreases from its initial
value of $\langle\Sigma(r_p,0)\rangle_\phi
\simeq9.39\times10^{-2}\ {\rm g\,cm^{-2}}$ to
$5.52\times10^{-3}$, $7.03\times10^{-4}$, and
$2.44\times10^{-4}\ {\rm g\,cm^{-2}}$ for the $2$, $4$, and
$8\,M_{\rm Jup}$ models, respectively. These values correspond to local
depletion factors of approximately $17$, $134$, and $385$.

\subsection{Vertical structure and three-dimensional accretion flows}
\label{subsec:cpd_vertial_structure}

In order to characterize how the circumplanetary region is fed and how its structure changes with planetary mass, we analyze the gas distribution in meridional slices, the angular geometry of the mass exchange across spherical surfaces, and the three-dimensional morphology of the flow. The detailed characterization of the accretion geometry focuses on the fiducial $4\,M_{\rm Jup}$ simulation during the interval of strongest accretion, identified from the temporal evolution of the mass budget discussed below.

\subsubsection{Vertical structure of the gas in meridional slices}
\label{subsubsec:cpd_meridional_structure_slice}

\begin{figure*}[t!]
\centering
\includegraphics[width=\textwidth]{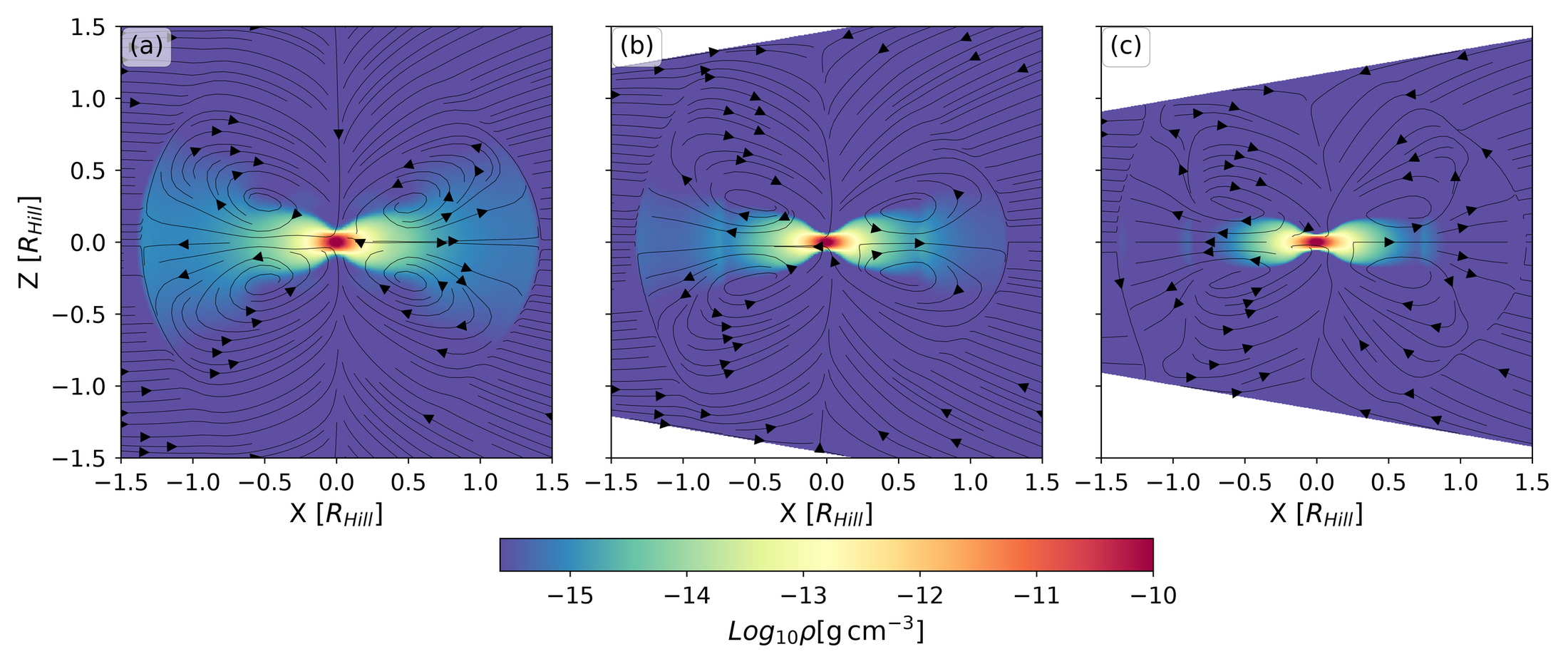}
\caption{Slices in the meridional $XZ$ plane of the gas volume density
(logarithmic scale) and instantaneous streamlines of the velocity
field projected onto the $XZ$ plane for planets of (a) $2\,M_{\rm Jup}$,
(b) $4\,M_{\rm Jup}$, and (c) $8\,M_{\rm Jup}$. All distances are expressed
in units of the planet's Hill radius. The star is located in the direction
of the negative $X$-axis.}
\label{fig:xz_streamlines_mass}
\end{figure*}

\begin{figure*}[t!]
\centering
\includegraphics[width=\textwidth]{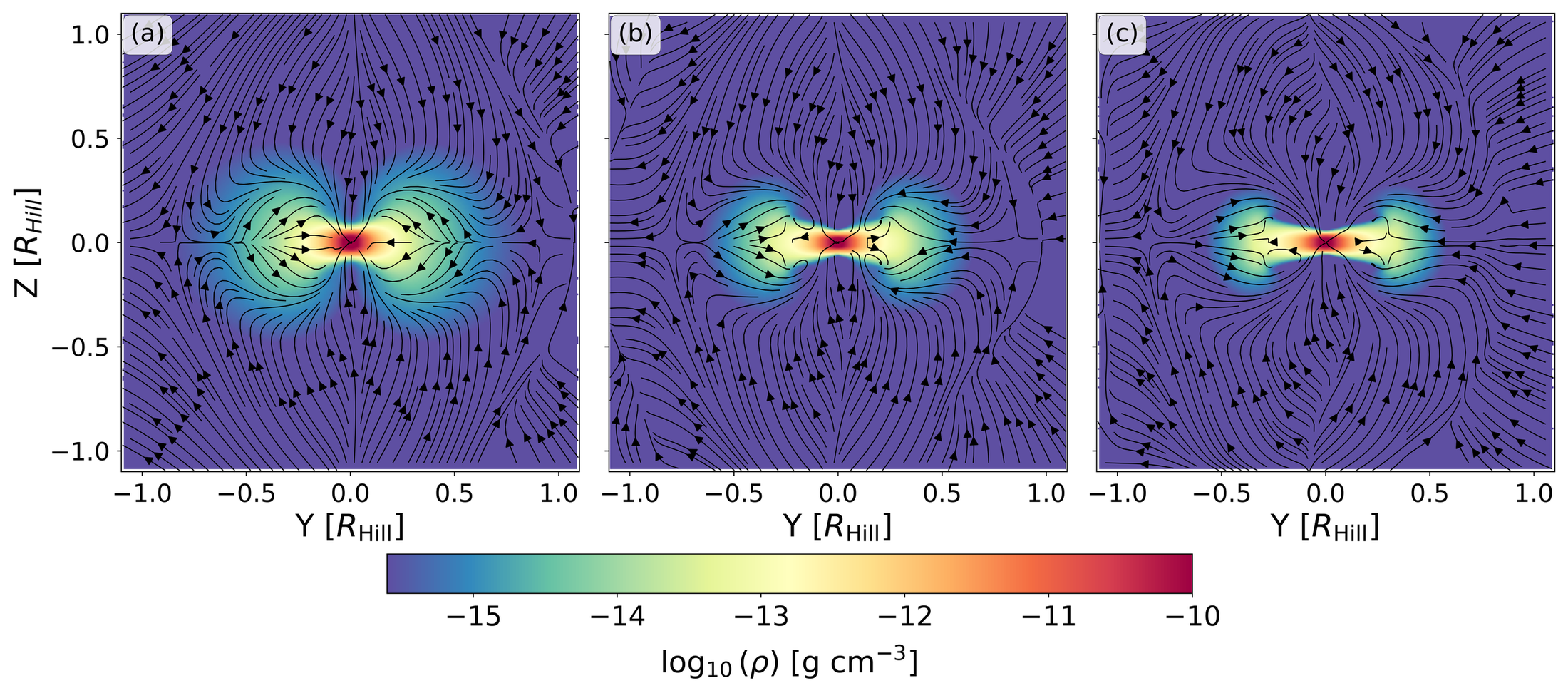}
\caption{Slices in the meridional $YZ$ plane of the gas volume density
(logarithmic scale) and instantaneous streamlines of the velocity
field projected onto the $YZ$ plane for planets of (a) $2\,M_{\rm Jup}$,
(b) $4\,M_{\rm Jup}$, and (c) $8\,M_{\rm Jup}$. All distances are expressed
in units of the planet's Hill radius.}
\label{fig:yz_streamlines_mass}
\end{figure*}

\autoref{fig:xz_streamlines_mass} and
\autoref{fig:yz_streamlines_mass} show the gas density and
instantaneous velocity field projected onto the $XZ$ and $YZ$
planes, respectively. The streamlines identify in-plane circulation and
inflow patterns, but do not imply closed three-dimensional trajectories.
The full three-dimensional flow geometry is examined separately in
\autoref{fig:stream3d}. In all cases, gas enters the Hill sphere from high
latitudes and interacts with the denser midplane region.

For $M_{\rm p}=2\,M_{\rm Jup}$, the gas within the Hill sphere remains vertically extended, with substantial density distributed over a large fraction of $|z|/R_{\rm Hill}$. The projected meridional velocity field exhibits broad circulation
patterns linking the upper disk layers to the planetary vicinity. In this regime, the imposed thermal support remains comparable to the depth of the planetary potential, limiting vertical confinement.

For $M_{\rm p}=4\,M_{\rm Jup}$, the gas becomes more concentrated toward the midplane, and the inflow is redirected through more clearly defined channels. Projected circulation remains visible, but the circumplanetary
region is more stratified and spatially organized.

For $M_{\rm p}=8\,M_{\rm Jup}$, the high-density gas is further compressed toward the midplane, and the vertical inflow is confined to narrower regions. The increase in planetary mass therefore drives a clear transition from a vertically extended envelope-like configuration to a more flattened and dynamically confined circumplanetary structure.

\subsubsection{Three-dimensional geometry of mass exchange}
\label{subsubsec:cpd_accretion_maps}

\begin{figure}
\centering
\includegraphics[width=\linewidth]{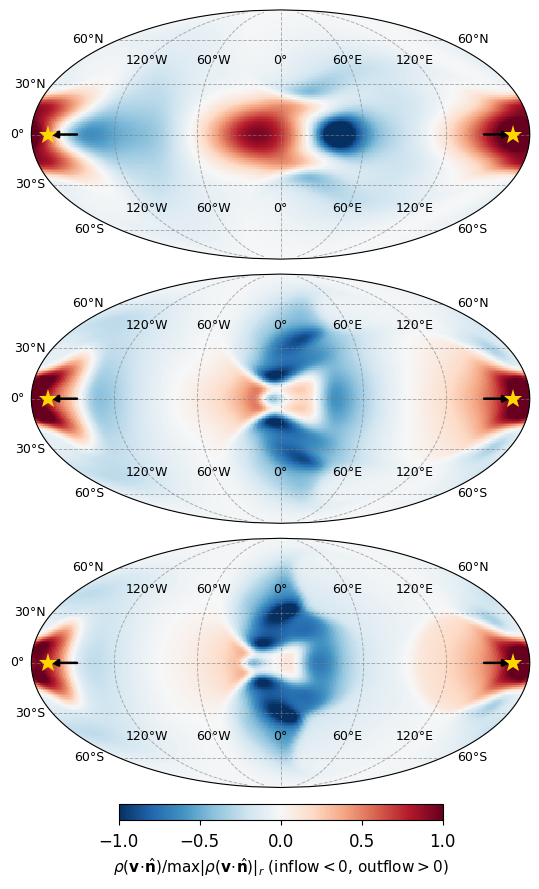}
\caption{Mollweide maps of the normalized mass flux,
$\rho(\mathbf{v}\cdot\hat{\mathbf{n}})/\max|\rho(\mathbf{v}\cdot\hat{\mathbf{n}})|_r$,
through spheres centered on the planet for the fiducial $4\,M_{\rm Jup}$ simulation at snapshot 50. The three panels correspond, from top to bottom, to radii $r=0.3$, $0.5$, and $0.6\,R_{\rm Hill}$. The maps highlight the angular geometry of inflow (negative values) and
outflow (positive values). The direction of the star is indicated with a star symbol and black arrows; by construction, longitudes near $0^\circ$ correspond to directions diametrically opposed to the planet-star line, while longitudes near $\pm 180^\circ$ point toward the star.}
\label{fig:mollweide_flux_norm}
\end{figure}

To characterize the angular geometry of the mass exchange, \autoref{fig:mollweide_flux_norm} shows Mollweide maps of the normalized mass flux through spherical surfaces centered on the planet in the fiducial $4\,M_{\rm Jup}$ simulation. The quantity displayed is $\rho\,(\mathbf{v}\cdot\hat{\mathbf{n}})$, normalized by its maximum absolute value at each radius, so that the maps emphasize the geometry of inflow and outflow independently of their absolute amplitude.

At $r=0.3\,R_{\rm Hill}$, the pattern is relatively simple and dominated by inflow concentrated near low latitudes, close to the midplane. Outflow occupies a smaller fraction of the surface and is angularly separated from the inflow regions. This behavior is consistent with the meridional slices, which show that the denser gas bound to the circumplanetary structure is concentrated toward the midplane.

At $r=0.5\,R_{\rm Hill}$, the angular distribution becomes more complex. Inflow extends to intermediate and high latitudes, while outflow appears in more clearly localized sectors. At this radius, the control surface intersects both the gas circulating within the circumplanetary region and the broader three-dimensional streams linked to the circumstellar disk.

At $r=0.6\,R_{\rm Hill}$, the anisotropy persists, but the inflow covers a larger fraction of the sphere, whereas the outflow is restricted to narrower regions. The mass exchange at this radius therefore reflects the transition between the inner circumplanetary flow and the surrounding disk more directly than at smaller radii.

The integrated fluxes at snapshot 50 support this picture. The absolute inflow and outflow rates decrease radially outwards: $\dot{M}_{\rm in}$ drops from $\simeq 8.9\times10^{-6}$ at $r=0.3\,R_{\rm Hill}$ to $2.4\times10^{-6}\,M_{\rm Jup}\,\mathrm{yr^{-1}}$ at $r=0.6\,R_{\rm Hill}$, with a similar decline observed for $\dot{M}_{\rm out}$. However, the net mass flux remains remarkably constant across all radii at $\dot{M}_{\rm net}\simeq -1.4$ to $-1.5\times10^{-6}\,M_{\rm Jup}\,\mathrm{yr^{-1}}$. The negative net flux indicates ongoing accretion at this epoch, while the radial decline of the absolute exchange rates confirms that the outer layers of the Hill sphere are increasingly dominated by recirculating flows rather than feeding the compact bound structure.

\subsubsection{Three-dimensional morphology of the gas}
\label{subsubsec:cpd_isodens_3d}

\begin{figure}
\centering
\includegraphics[width=1.02\columnwidth]{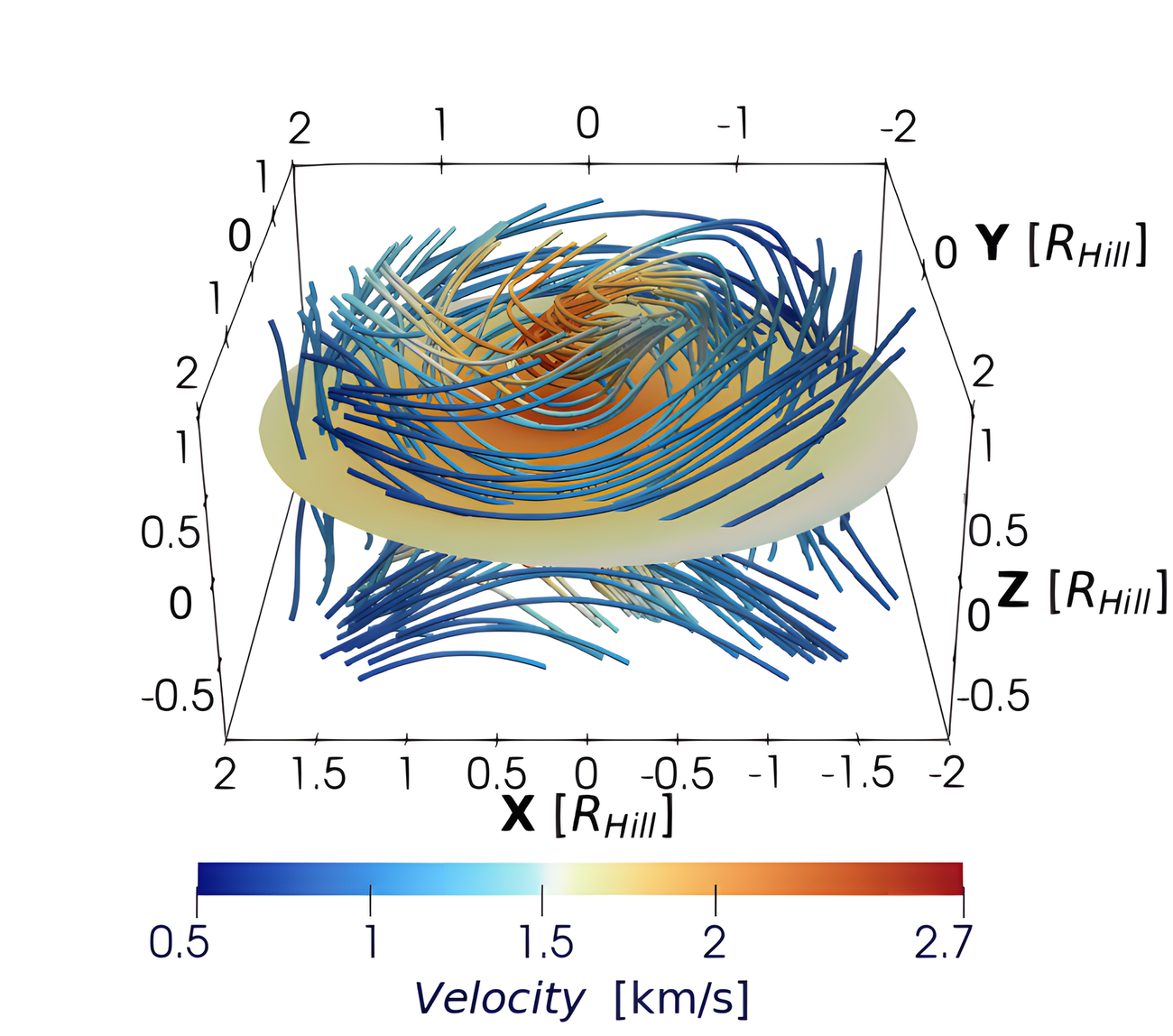}
\caption{
Three-dimensional structure of the gas flow in the circumplanetary environment, shown by means of streamlines in a reference frame centered on the planet. The streamlines are plotted in units of the Hill radius, while the
color indicates the magnitude of the total gas velocity. The shaded plane marks the orbital midplane ($z=0$). The figure illustrates the connection between meridional flows coming from the upper layers of the circumstellar disk and the rotational flow that is established in the inner region of the CPD. These visualizations were obtained with \textsc{ParaView} from the original, non-interpolated data.}
\label{fig:stream3d}
\end{figure}

\begin{figure*}[t!]
\centering
\includegraphics[width=1.05\textwidth]{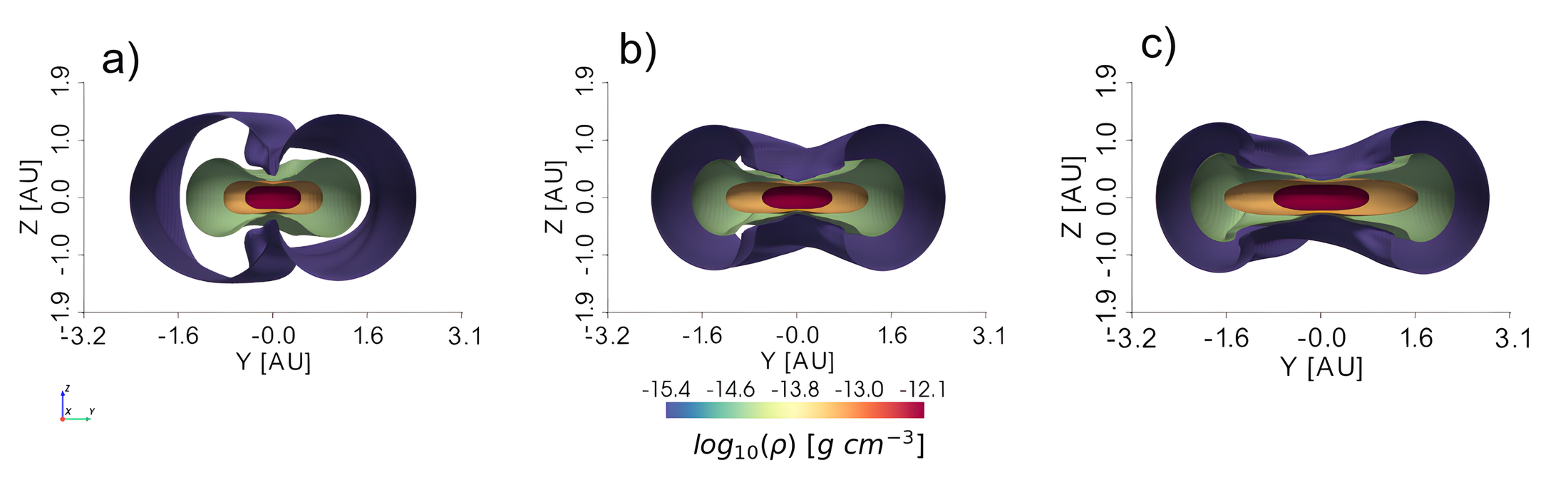}
\caption{Three-dimensional gas density isosurfaces in the circumplanetary environment, visualized with \texttt{FARGOpy}. The same density values are used in all panels, enabling a direct comparison of the three-dimensional gas morphology for different planetary masses: (a) $2\,M_{\rm Jup}$, (b) $4\,M_{\rm Jup}$, and (c) $8\,M_{\rm Jup}$. Distances are expressed in physical units (AU), and the midplane corresponds to $Z=0$.}
\label{fig:isosurfaces_3d_density}
\end{figure*}

\autoref{fig:stream3d} provides a three-dimensional view of the flow topology. The streamlines show that accretion onto the circumplanetary region is not planar or isotropic, but is instead dominated by descending meridional flows that are redirected toward the midplane. Within the inner fraction of the Hill sphere, the streamlines bend
progressively toward the orbital plane, indicating the establishment of a rotationally organized flow. Farther out, the larger spread in both angle and height reflects the coexistence of inflow, outflow, and large-scale circulation.

The same trend is visible in the three-dimensional density isosurfaces shown in \autoref{fig:isosurfaces_3d_density}. In all three models, the highest-density gas is concentrated in a flattened region around the midplane, while lower-density isosurfaces delineate a more extended envelope connected to the circumstellar disk. As the planetary mass increases, the gas becomes more confined toward the midplane and occupies a larger fraction of the Hill sphere radially. The lowest-mass model retains a more rounded outer morphology, whereas the intermediate- and high-mass cases show progressively flatter inner structures, consistent with a more developed circumplanetary disk.

Taken together, the meridional slices, the angular flux maps, and the three-dimensional visualizations all support the same picture: the circumplanetary gas is fed by intrinsically three-dimensional accretion and organizes into a denser inner component surrounded by a lower-density, recirculating envelope whose vertical extent decreases with increasing planetary mass.

\subsection{Enclosed mass evolution and net mass-change rates}
\label{subsec:cpd_accretion}

\begin{figure*}
\centering
\includegraphics[width=\textwidth]{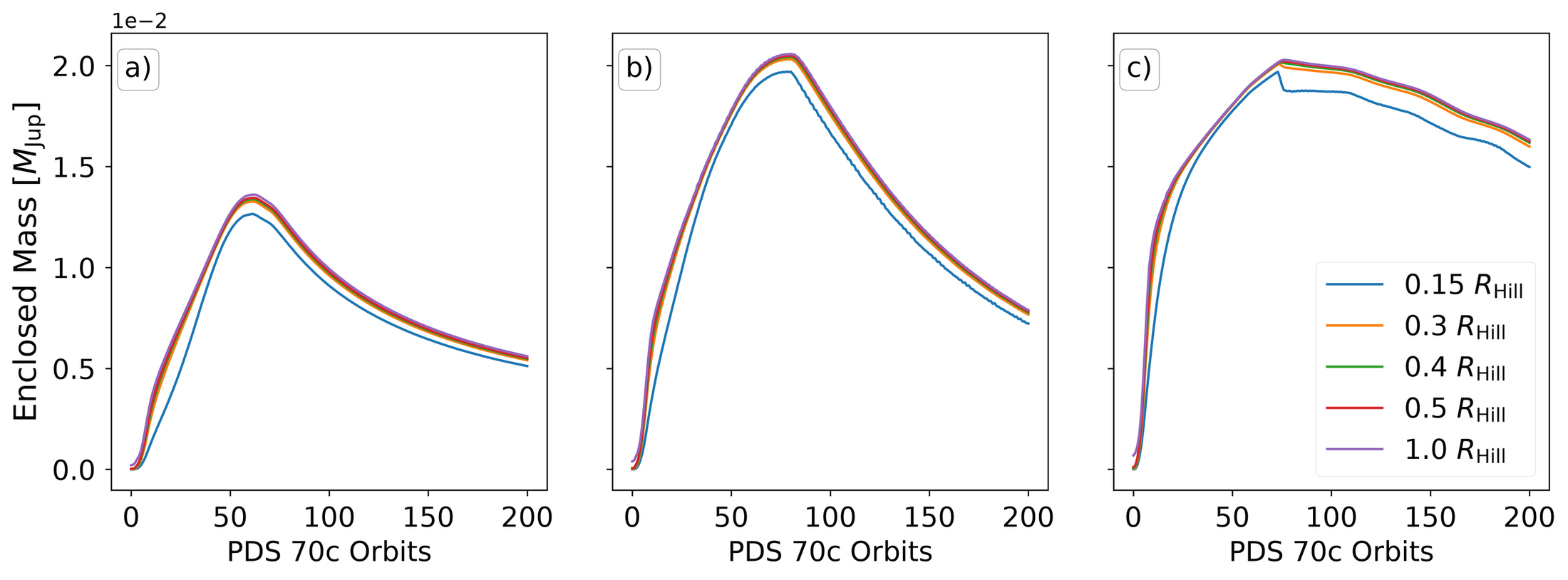}
\caption{
Time evolution of the enclosed mass $M(<R,t)$ within spheres centered on the planet, for the three planetary masses considered and different integration radii $R$. Distances are expressed in units of the Hill radius.}
\label{fig:mass_time}
\end{figure*}

\begin{figure*}
\centering
\includegraphics[width=0.9\textwidth]{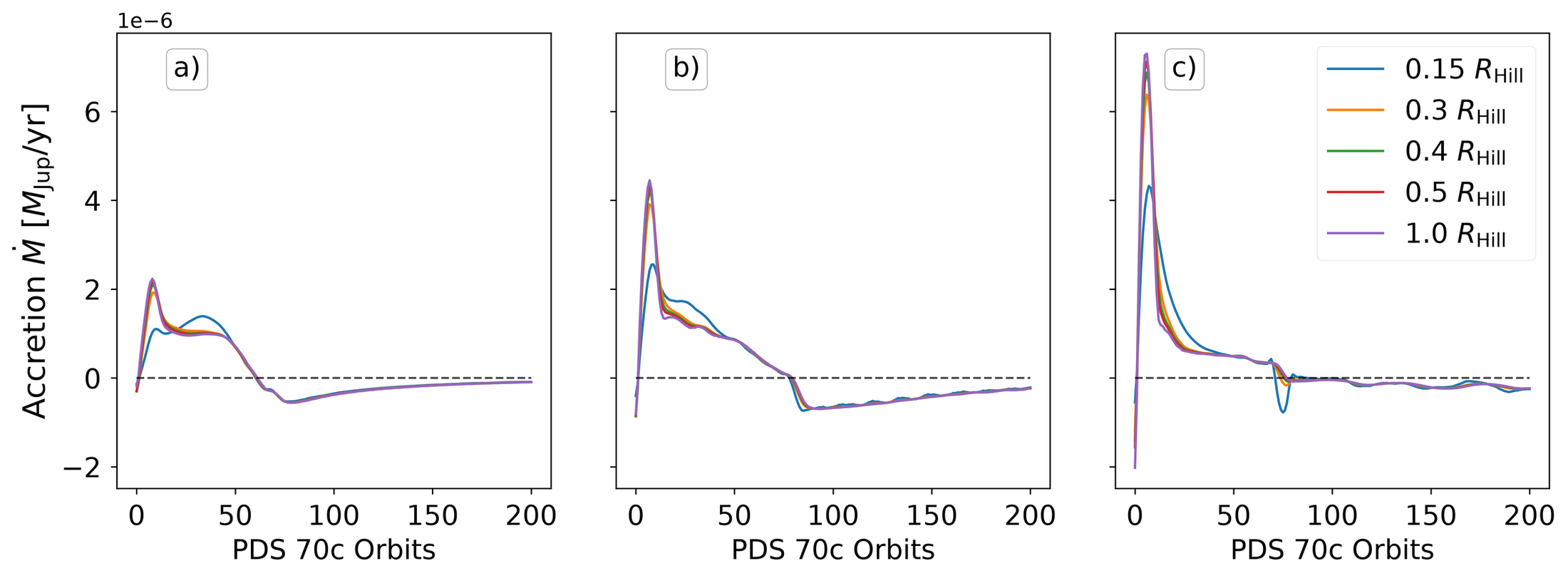}
\caption{Time evolution of the net mass-change rate
$\dot{M}(<R,t)$, defined as the time derivative of the mass enclosed within
a sphere centered on the planet, for different integration radii $R$.
Positive and negative values indicate increasing and decreasing
enclosed mass, respectively. The three panels correspond to the three
planetary masses considered in this work. The time axis is expressed in
orbits of PDS~70c.}
\label{fig:mdot_time}
\end{figure*}

To quantify the temporal evolution of the circumplanetary reservoir, we compute the gas mass enclosed within spheres centered on the planet.
\begin{equation}
M(<R,t)=\int_{V(R)} \rho(\mathbf{x},t)\,{\rm d}V,
\end{equation}
where $V(R)$ is a sphere of radius $R=f\,R_{\rm Hill}$. From this quantity, we define the net mass-change rate
\begin{equation}
\dot{M}(<R,t)=\frac{{\rm d}}{{\rm d}t}M(<R,t),
\end{equation}
after temporal smoothing to suppress short-period numerical fluctuations. Late-time values are summarized by the median over the last 30\%
of the simulation, with the 16th and 84th percentiles used to characterize
their variability. These statistics describe the late-time evolution and do
not imply that the enclosed mass has reached a steady state.

The enclosed mass, shown in \autoref{fig:mass_time}, rises rapidly
during the first tens of orbits, reaches a maximum between approximately
$50$ and $80$ orbits, and subsequently decreases throughout the remainder
of the simulations. The initial increase reflects the adjustment of the gas
to the growing planetary potential and the evolving smoothing length. After
approximately $100$ orbits, when the smoothing transition is complete, the
mass loss becomes slower but remains systematic. Because no gas is removed
at the planetary position, this decrease does not represent accretion onto
the planet; instead, it reflects net outward transport across the boundaries
of the planet-centered control volumes and redistribution into the surrounding
disk.

A key result is that the enclosed mass depends only weakly on the integration radius once $R$ exceeds a few tenths of the Hill radius. Using medians over the last 30\% of the runs, the increase from $0.15\,R_{\rm Hill}$ to $1.0\,R_{\rm Hill}$ is only $\sim 7$--$10$\%: for $2\,M_{\rm Jup}$, the mass increases from $\simeq 5.8\times10^{-3}$ to $\simeq 6.4\times10^{-3}\,M_{\rm Jup}$; for $4\,M_{\rm Jup}$, from $\simeq 9.1\times10^{-3}$ to $\simeq 9.9\times10^{-3}\,M_{\rm Jup}$; and for $8\,M_{\rm Jup}$, from $\simeq 1.64\times10^{-2}$ to $\simeq 1.76\times10^{-2}\,M_{\rm Jup}$. Most of the dense circumplanetary gas is therefore concentrated in the inner region, while the outer Hill sphere contributes mainly a low-density, recirculating component.

The corresponding net mass-change rates are shown in
\autoref{fig:mdot_time}. The initial evolution contains strong transients
associated with the introduction of the planetary potential. After the
enclosed mass reaches its maximum, $\dot{M}(<R,t)$ remains predominantly
negative, consistently with the gradual decline shown in
\autoref{fig:mass_time}. Its magnitude decreases with time but does not
vanish completely. The late-time
behavior is described as a slowly declining, recirculating reservoir,
rather than as a quasi-steady or monotonically growing CPD.

\subsection{Shock formation and dynamical dissipation within the CPD}
\label{subsec:cpd_shocks}

\begin{figure*}[t!]
\centering
\includegraphics[width=\textwidth]{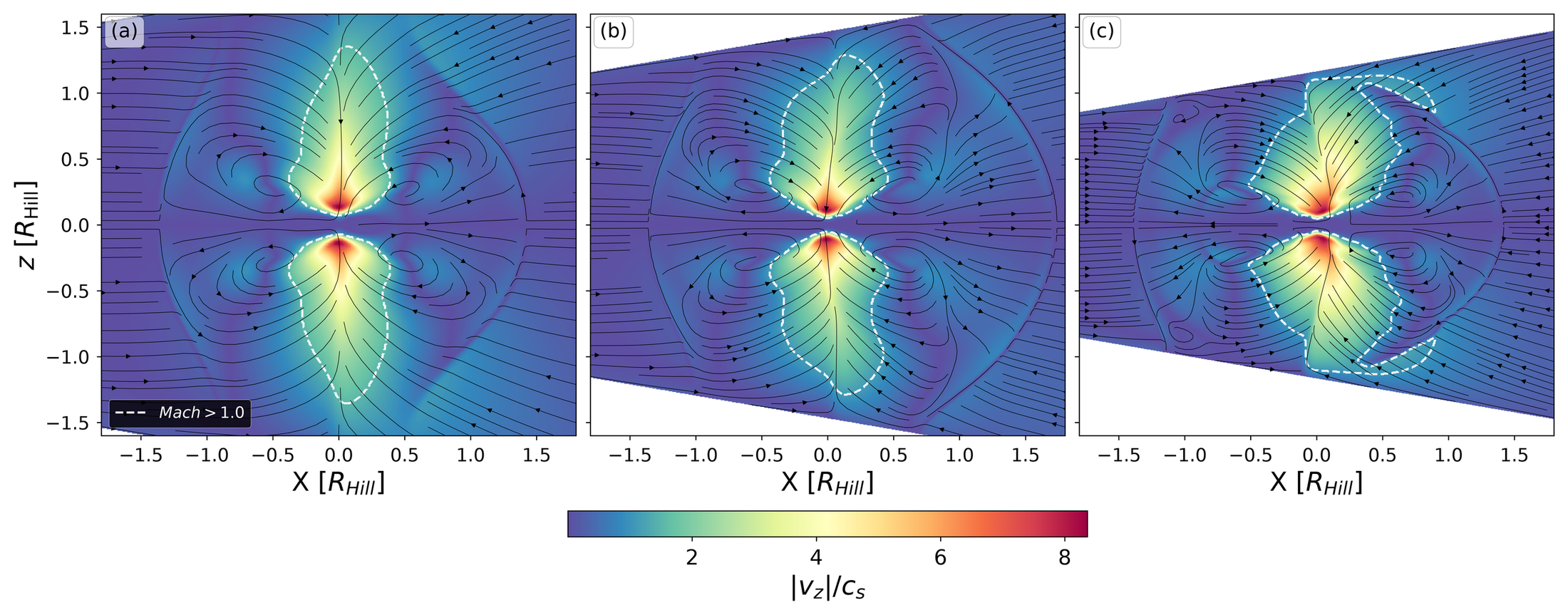}
\caption{Map of the normalized vertical velocity $\mathcal{M}=|v_z|/c_s$ (Mach number) in the meridional $XZ$ plane, with streamlines overplotted and the dashed contour indicating supersonic regions ($\mathcal{M}>1$), for (a) $2\,M_{\rm Jup}$, (b) $4\,M_{\rm Jup}$, and (c) $8\,M_{\rm Jup}$. All distances are expressed in units of the planet's Hill radius. The star is located in the direction of the negative $X$-axis.}
\label{fig:mach_xz}
\end{figure*}

\begin{figure*}
\centering
\includegraphics[width=0.9\textwidth]{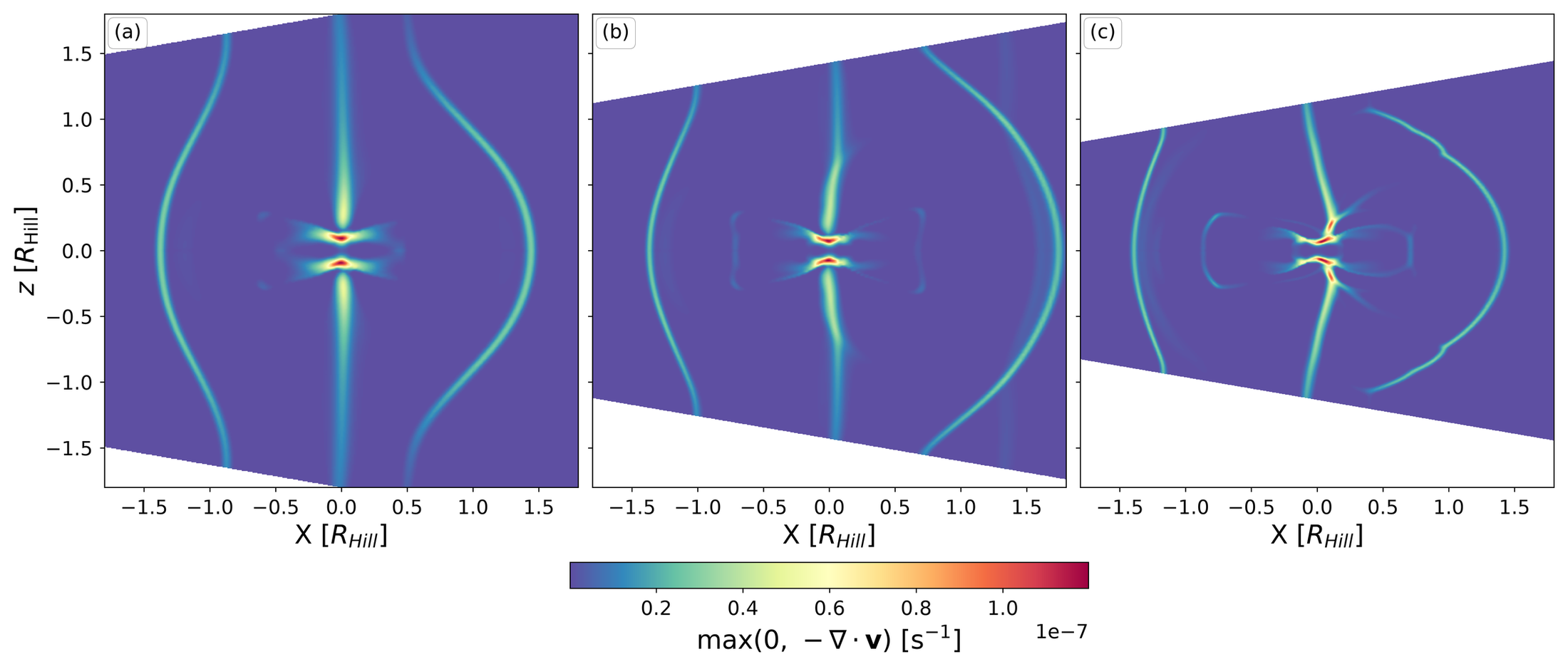}
\caption{
Compressibility map in the meridional $XZ$ plane, defined as $\mathcal{C}=\max(0,-\nabla\cdot\mathbf{v})$, for planetary masses of (a) $2\,M_{\rm Jup}$, (b) $4\,M_{\rm Jup}$, and (c) $8\,M_{\rm Jup}$. The tracer selects regions of converging flow and has units of ${\rm s^{-1}}$. All distances are expressed in units of the planetary Hill radius. The star is located in the direction of the negative $X$ axis.}
\label{fig:compressibility_xz}
\end{figure*}

The three-dimensional accretion geometry implies that gas enters the Hill sphere through high-latitude channels and can be accelerated to supersonic velocities before interacting with the denser circumplanetary region. To characterize where this interaction is dynamically strongest, we analyze both the vertical Mach number, $\mathcal{M}_Z$, and the compression tracer, $\mathcal{C}$ (see \autoref{eq:compression_tracer}), in the $XZ$ plane.

The supersonic regions traced by $\mathcal{M}_Z$ in \autoref{fig:mach_xz} are vertically extended and remain confined to relatively narrow radial regions. In all three models, gas with $\mathcal{M}_Z>1$ reaches heights on the order of $|z|\sim 1.1$--$1.35\,R_{\rm H}$ while remaining concentrated within $|x|\lesssim 0.4$--$0.5\,R_{\rm H}$. These structures, therefore, trace vertically extended inflow channels rather than a globally supersonic circumplanetary flow.

The regions of strongest compression shown in \autoref{fig:compressibility_xz} are much more localized. Their highest values are confined to compact regions close to the planet, typically within $|x|\lesssim 0.1$--$0.15\,R_{\rm H}$ and $|z|\lesssim 0.4\,R_{\rm H}$, with the most intense compression concentrated in lobes near $|z|\sim 0.08$--$0.15\,R_{\rm H}$. This spatial separation demonstrates that not all supersonic gas produces strong compression; rather, dissipation is concentrated where the vertical inflow decelerates and couples to the denser circumplanetary flow.

More extended arcs of enhanced compression are also present farther from the planet and align broadly with the large-scale circulation within the Hill sphere. Their coexistence with the supersonic inflow channels supports a picture in which shocks are localized to specific interaction layers rather than distributed throughout the entire flow. The $4\,M_{\rm Jup}$ model exhibits the cleanest and most symmetric morphology, whereas the $8\,M_{\rm Jup}$ case shows stronger but more irregular structures, indicating that the degree of organization depends not only on the depth of the potential well but also on the redistribution of mass and momentum within the Hill sphere.

Supersonic vertical inflows are not unique to our simulations; for example,
\citet{fung-2019} find vertical inflow velocities reaching approximately
$4$-$5\,c_s$ in three-dimensional isothermal CPD simulations.
Because the simulations are locally isothermal and do not evolve the energy
equation, these quantities should be interpreted as identifying where
dissipation would be most efficient under more realistic thermodynamic
conditions rather than as direct predictions of the thermal response. Even so,
the combination of \autoref{fig:mach_xz} and
\autoref{fig:compressibility_xz} shows clearly that shock formation in the
circumplanetary environment is highly localized and is controlled by the
geometry of the three-dimensional accretion flow.

\subsection{Kinematics of the circumplanetary disk}
\label{subsec:cpd_kinematics_curve}

The kinematical properties of the circumplanetary gas determine whether the material captured within the Hill sphere settles into a rotationally supported disk-like configuration or remains predominantly sustained by pressure support and transient three-dimensional flows. We therefore analyze both azimuthally averaged radial profiles of the normalized azimuthal velocity, $\langle v_\phi/V_{K,p}\rangle(r)$, and two-dimensional maps of $|v_\phi|/V_{K,p}$ in the midplane. Here, $V_{K,p}$ is the circular velocity associated with the softened planetary potential used in the simulations. The velocity profiles and maps are evaluated after 200 orbits, when the smoothing transition is complete and the evolution of the enclosed mass has slowed.

\begin{figure}
\centering
\includegraphics[width=\columnwidth]{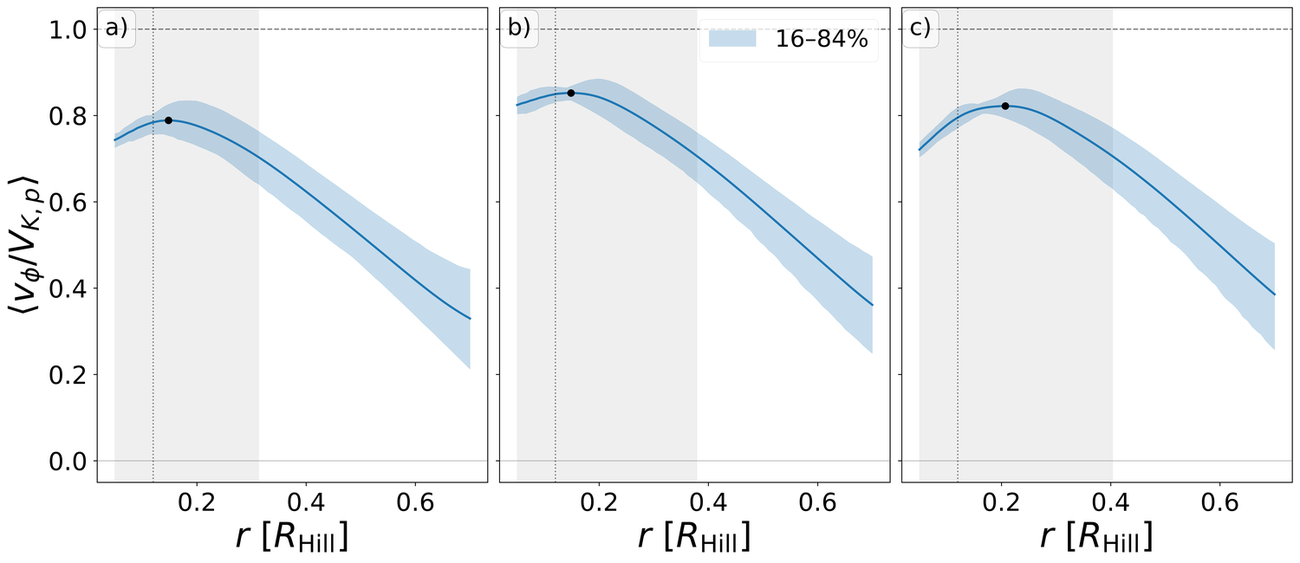}
\caption{
Azimuthally averaged radial profiles of the normalized azimuthal velocity
$\langle v_\phi/V_{K,p}\rangle(r)$ in the midplane, for (a) $2\,M_{\rm Jup}$,
(b) $4\,M_{\rm Jup}$, and (c) $8\,M_{\rm Jup}$, measured after 200 orbits.
Solid lines show the azimuthal average, while blue shaded regions indicate
the 16--84\% interval, quantifying the level of non-axisymmetry. Gray shaded
regions mark the radial interval satisfying
$0.7\leq\langle v_\phi/V_{K,p}\rangle<1$, used to define the CPD radius
$R_{\rm CPD}$. The vertical dotted line marks the final
smoothing length, $\epsilon_{\rm fin}=S_{\rm fin}R_{\rm Hill}=0.12\,R_{\rm Hill}$.
}
\label{fig:rotcurve}
\end{figure}

\begin{figure*}
\centering
\includegraphics[width=\textwidth]{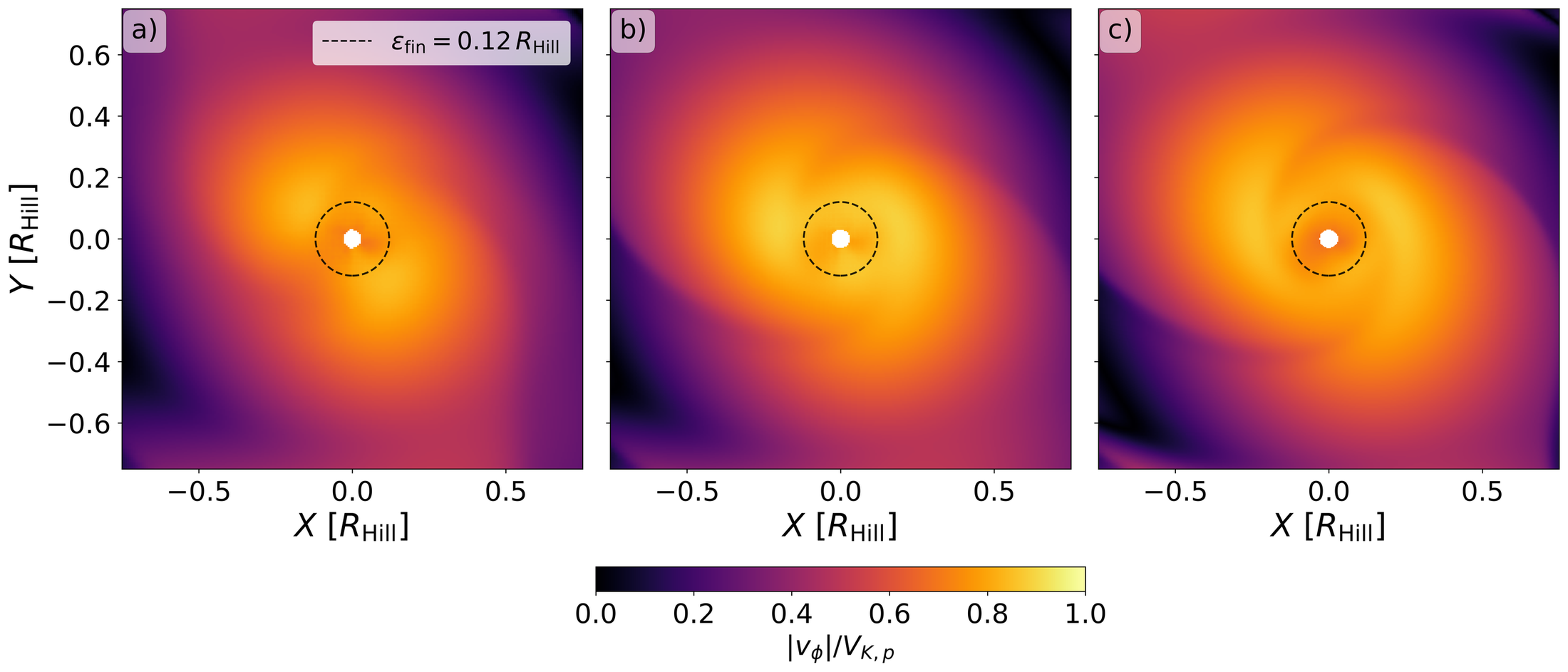}
\caption{
Maps of the normalized planetocentric azimuthal velocity $|v_\phi|/V_{K,p}$ in the midplane for (a) $2\,M_{\rm Jup}$, (b) $4\,M_{\rm Jup}$, and (c) $8\,M_{\rm Jup}$, measured after 200 orbits. The dashed circle marks the smoothing length, $\epsilon=0.12\,R_{\rm Hill}$, separating the innermost softened region from the resolved outer flow. Outside this radius, the maps show an inner rotationally organized region and an outer region dominated by spiral asymmetries and interaction with the circumstellar disk. Distances are expressed in units of the Hill radius.}
\label{fig:rotmap}
\end{figure*}

\autoref{fig:rotcurve} shows that all three models develop an inner region of organized rotation, but this rotation remains globally sub-Keplerian. We define the CPD radius, $R_{\rm CPD}$, as the outermost radius of the
prograde, sub-Keplerian, rotationally supported region satisfying
$0.7\leq\langle v_\phi/V_{K,p}\rangle<1$.
For planetary masses of $2$, $4$, and $8\,M_{\rm Jup}$, we obtain
$R_{\rm CPD}\simeq0.31$, $0.38$, and $0.40\,R_{\rm Hill}$, respectively.
The corresponding maxima of
$\langle v_\phi/V_{K,p}\rangle$ are $0.789$, $0.852$, and $0.822$,
located at $r\simeq0.148$, $0.148$, and $0.207\,R_{\rm Hill}$,
respectively.

The corresponding maps in \autoref{fig:rotmap} show that the region outside the smoothing length contains a coherent inner rotational pattern, while the outer flow exhibits strong spiral asymmetries. These non-axisymmetric structures arise naturally in a CPD fed by three-dimensional accretion, where gas enters the Hill sphere through preferential channels and redistributes its angular momentum before partially settling into orbit around the planet. Their prominence increases with planetary mass, reflecting the stronger perturbation exerted by more massive planets and the stronger coupling between the inner rotationally organized region and the outer transition zone.

The normalized radial profiles remain broadly similar across the explored mass range, suggesting that the Hill radius sets the dominant dynamical scale of the system. The inferred CPD radius increases from
$R_{\rm CPD}\simeq0.31$ to $0.40\,R_{\rm Hill}$ from the lowest- to the
highest-mass model. The circumplanetary gas is therefore rotationally organized in the inner region, but the flow remains globally sub-Keplerian and dynamically connected to the surrounding circumstellar disk.

\section{Discussion}
\label{sec:discussion}

\subsection{Comparison with previous circumplanetary-flow models}
\label{subsec:discussion_previous}

The flow morphology obtained here is consistent with previous
three-dimensional studies in which circumplanetary regions are supplied by
meridional inflows and remain connected to the circumstellar disk through
inflow, outflow, and recirculation
\citep{Tanigawa-2012,Szulagyi-2016,li-2023}. Our results add a quantitative
characterization of this exchange for a PDS~70c-motivated setup, including
angularly resolved fluxes, enclosed-mass profiles, compression maps, and
planet-centered rotational velocities.

The increasing confinement and rotational organization with planetary
mass are also consistent with the near-isothermal survey of
\citet{Sagynbayeva2025}. For the adopted aspect ratio, all three planets lie
above the local thermal mass, and our simulations show that the circumplanetary
structure continues to become denser and more flattened between
$2$ and $8\,M_{\rm Jup}$. The inferred CPD radius is approximately
$0.31$--$0.40\,R_{\rm Hill}$, a scale comparable to the radial
extent of rotating circumplanetary flows found in other global simulations,
although the precise value depends on the thermodynamic treatment and on the
criterion used to define the CPD boundary
\citep{fung-2019,li-2023}.

The gas nevertheless remains substantially sub-Keple-rian, reaching
$\langle v_\phi/V_{K,p}\rangle\simeq0.79$ - $0.85$. Previous comparisons between isothermal,
adiabatic, and radiative calculations show that the thermal evolution strongly
affects whether the gas forms a disk or a pressure-supported envelope
\citep{Szulagyi-2016,fung-2019, krapp-2024}. Because all three models
adopt the same locally isothermal closure, we cannot quantify how much of the
central density enhancement is caused by the thermodynamic prescription itself.
The prescribed temperature may nevertheless favor stronger central
concentration by preventing compressional and shock heating from increasing
the local pressure support. In particular, the sub-Keplerian
rotation in our models should not be interpreted as a direct consequence of the enhanced central density in the isothermal treatment, but rather as the
combined result of pressure support and the three-dimensional, non-circular flow within the Hill sphere. Our results therefore describe the rotational
structure obtained in a prescribed-temperature, efficient-cooling limit rather than a universal CPD state.

\subsection{Implications for PDS~70c}
\label{subsec:discussion_pds70c}

For the adopted stellar mass and orbital radius, the Hill radius
ranges from approximately $3.2$ to $5.1$ au across the explored planetary
masses. The inferred CPD radius is approximately
$0.31$--$0.40\,R_{\rm Hill}$, corresponding to physical radii of about
$1.0$--$2.0$ au. These gas-dynamical scales are larger than the compact
continuum scale inferred around PDS~70c
\citep{isella-2019,benisty-2021}. However, the comparison is not direct:
the simulations characterize the radial extent of organized gas rotation,
whereas the millimeter emission traces dust whose distribution also depends
on temperature, opacity, grain evolution, and radial drift
\citep{portilla-revelo-2021,shibaike-2024}.

In this context, \citet{portilla-revelo-2023} infer a gas surface density of $\Sigma\simeq7\times10^{-3}\ {\rm g\,cm^{-2}}$ at the orbital location of PDS~70c. The value obtained in our $2\,M_{\rm Jup}$ model, $5.52\times10^{-3}\ {\rm g\,cm^{-2}}$, is comparable to this observationally motivated estimate, whereas the $4$ and $8\,M_{\rm Jup}$
models produce substantially deeper gaps.

The simulations indicate that the compact inner structure is not an
isolated reservoir. Most of the enclosed gas is concentrated within a few
tenths of the Hill radius, while the outer region is crossed continuously by
anisotropic inflow, outflow, and recirculation. This provides a hydrodynamic
interpretation in which circumplanetary material around PDS~70c remains
dynamically connected to the gas within the disk cavity, consistent with the
presence of larger-scale stream-like structures inferred observationally and
numerically \citep{toci-2020,christiaens-2024}.

\subsection{Physical scope and methodological contribution}
\label{subsec:discussion_scope}

The locally isothermal approximation provides a controlled
efficient-cooling limit in which the thermal structure is prescribed, allowing
the three-dimensional organization of the flow to be isolated from additional
thermodynamic complexity. Within this framework, the simulations reliably characterize the dependence of the circumplanetary morphology, mass exchange,
and rotational structure on planetary mass. Quantities that depend directly on
the thermal response of the gas, such as post-shock temperatures, radiative
emission, and dust evolution, require a more complete thermodynamic treatment.

The models therefore provide a hydrodynamic baseline for
PDS~70c-motivated disk conditions. They quantify the three-dimensional flow
geometry, characteristic spatial scales, mass redistribution, and rotational
state, and establish a reference for future calculations including radiative
thermodynamics and dust.

A complementary outcome is the development and application of
\fargopy\ as a reproducible framework for analyzing three-dimensional
\fargotd\ outputs \citep{MurilloGonzalez2026}. The package combines Cartesian
slices, planet-centered coordinate transformations, fluxes through closed
surfaces, enclosed-mass measurements, and kinematic profiles within a common
workflow. The present simulations therefore serve both as a physical study of
a PDS~70c-like system and as a demonstration of analysis tools that can be
applied consistently to future models with additional thermodynamic and dust
physics.

\section{Conclusions}
\label{sec:conclusions}

We presented three-dimensional hydrodynamical simulations of a
PDS~70c-motivated system for planetary masses of $2$, $4$, and
$8\,M_{\rm Jup}$. Increasing planetary mass produces a progressively denser,
flatter, and more rotationally organized circumplanetary structure. The CPD radius is
$R_{\rm CPD}\simeq0.31$--$0.40\,R_{\rm Hill}$, while the inner gas remains
sub-Keplerian, with $v_\phi/V_{K,p}\sim0.79$--$0.85$.

The circumplanetary region is supplied through intrinsically
three-dimensional flows and remains connected to the circumstellar disk
through anisotropic inflow, outflow, and recirculation. Most of the enclosed
gas is concentrated within a few tenths of the Hill radius, and the strongest
compression occurs where descending flows interact with the dense inner
region.

These results establish an efficient-cooling hydrodynamic baseline
for PDS~70c rather than a prediction of its thermal or radiative appearance.
Together with the development of \fargopy, they provide both a quantitative
description of the simulated circumplanetary flow and a reusable framework
for analyzing future three-dimensional \fargotd\ models.

\section*{Data Availability}
The post-processing package \webfargopy, along with the \texttt{Jupyter} notebooks required to reproduce the analysis and figures presented in this paper, is publicly available at the \texttt{GitHub} repository of \fargopy,  \href{https://github.com/seap-udea/fargopy}{https://github.com/seap-udea/fargopy}. A representative sample of the medium-resolution simulation data is provided within the repository documentation.

\section*{Acknowledgments}
\label{sec:Acknowledgments}

We are grateful for the valuable comments and suggestions offered by the two anonymous reviewers. M.M. acknowledges financial support from FONDECYT Regular 1241818. This work was made possible thanks to the use of open-source software developed by the scientific community. In particular, we acknowledge the use of the hydrodynamic code \texttt{FARGO3D} \citep{Benitez-Llambay2016} and the analysis package {\texttt{FARGOpy}}, developed in the framework of this work. Data analysis and visualization were carried out using tools from the {\textsc Python} ecosystem, including {\textsc numpy} \citep{Numpy2020}, {\textsc scipy} \citep{Scipy2020}, {\textsc matplotlib} \citep{Matplotlib2007}, {\textsc pandas} \citep{Pandas2010}, {\textsc scikit-learn} \citep{pedregosaScikitlearnMachineLearning2011}, and {\textsc ParaView} \citep{Ahrens2005ParaView}.

\appendix
\section{FARGOpy in detail}
\label{app:fargopy}

In this appendix, we summarize some of the most important features of \webfargopy, provide examples of code developed with the package, and discuss the advantages it offers compared to the traditional approach used to process \fargotd\ outputs. Additionally, we present some of the validation tests used to ensure that the package performs one of the most important tasks for which it was designed, namely, interpolating fields on one-, two-, and three-dimensional grids. These appendices are not intended to be a complete presentation of the package. They are in fact an abridged version of the content of the technical report published in \citealt{MurilloGonzalez2026}\footnote{The thecnical report is publicly available in \href{https://github.com/seap-udea/fargopy/blob/main/science/introducing-fargopy/MurilloZuluagaMontesinos2026-IntroducingFARGOpy.pdf}{this url}.}

\subsection{General features}

\webfargopy\ is an open-source post-processing and visualization package for hydrodynamic simulations produced with \fargotd. It provides a reproducible analysis pipeline tailored to refined spherical grids, where curvilinear geometry and non-uniform spacing make generic tools error-prone.

The package is published and publicly available through the official Python Package Index (PyPI) repository, which allows direct installation via standard \texttt{Python} package managers. \webfargopy\ is distributed under the \emph{GNU Affero General Public License}, a free software license approved by the OSI, and is compatible with platform-independent operating systems. It falls within the areas of computational astronomy, fluid dynamics (CFD), and magnetohydrodynamics (MHD), with full support for \texttt{Python~3}.

\subsection{Workflow and data management}

\webfargopy\ organizes a simulation through the \texttt{Simulation} object, which (i) parses the run directory, (ii) loads snapshots and planet files, and (iii) exposes fields in physical units. The recommended workflow is:
\begin{enumerate}
    \item Instantiate a \texttt{Simulation} from an output directory (local or downloaded precomputed dataset).
    \item Select snapshots and load one or more fields (e.g. density, velocity components) with an explicit slice specification.
    \item Convert to a consistent unit system and perform resampling/interpolation as needed.
    \item Produce derived diagnostics (e.g. surface integrals, enclosed mass, fluxes) and figures.
\end{enumerate}

\noindent A minimal end-to-end example (tested with \webfargopy\ v1.2.X) is presented below\footnote{All code examples included in this manuscript can be accessed in the package's \texttt{GitHub} repository via \href{https://github.com/seap-udea/fargopy/blob/main/science/introducing-fargopy/notebooks/introducing-example-snippets.ipynb}{this link}.}

\begin{python}
# Tested in versions 1.2.X of FARGOpy
import fargopy as fp
import numpy as np
import matplotlib.pyplot as plt

path = fp.Simulation.download_precomputed('p3disoj')
sim = fp.Simulation(output_dir=path)
sim.units("CGS")

fields_cartesian = sim.load_field(
    fields='gasdens',
    snapshot=[1, 10],
    coords='cartesian',
    slice='theta=1.567' 
)

snapshot = 9
x = fields_cartesian.var1_mesh[snapshot]
y = fields_cartesian.var2_mesh[snapshot]

gasdens_plane = fields_cartesian.gasdens_mesh[snapshot]

plt.pcolormesh(x, y, gasdens_plane, cmap='Spectral_r')
\end{python}

For this example, we first downloaded one of the precomputed simulations \fargotd\ provided with the package. These precomputed simulations are intended to facilitate the learning process without the burden of waiting for a partial or full simulation of \fargotd\ (for a list of precomputed simulations, use the function \verb|fp.Simulation.list_precomputed()|).  Subsequently, a slice of the gas density field is loaded at a specific time (snapshot). The slice corresponds to a horizontal section in the middle of the disk ($\theta=\pi/2\approx 1.567$). You can copy this code snippet directly from \href{https://github.com/seap-udea/fargopy/blob/main/science/pds70c-isothermal/codes/Murillo_Zuluaga_Montesinos_PDS70c_CodeSnippets.ipynb}{this file}.

Anyone familiar with \fargotd\ can attest that the same procedure, programmed without the help of \fargopy, takes, depending on experience, anywhere from several tens of lines to more than a hundred lines of code.

\subsection{Multidimensional interpolation}
\label{app:fargopy_interpolation}
Many analysis tasks require mapping from the native spherical mesh to regular Cartesian or cylindrical grids (e.g., for midplane cuts, meridional sections, or spherical surfaces). \webfargopy\ therefore implements:
\begin{itemize}
    \item Consistent access to the native coordinates (cell centers/edges) and metric factors;
    \item Generic 2D and 3D interpolation wrappers with explicit control of the target grid;
    \item Utilities to validate round-trip interpolation (native \(\to\) resampled \(\to\) native).
\end{itemize}

\subsubsection{Implemented interpolation schemes}
For 2D resampling, the package provides bilinear/biquadratic schemes and (optionally) higher-order methods depending on the available backends. For 3D, interpolation is performed either directly in \((r,\theta,\phi)\) or after mapping to Cartesian coordinates, depending on the requested target geometry.

In the following snippet, we build a regular \((X,Z)\) mesh for a meridional slice and interpolate a scalar (density) and a vectorial (field):

\begin{python}
fields = sim.load_field(
    fields=['gasv','gasdens'],
    slice="r=[0.6,1.4], phi=0",
    snapshot=9
)
# Create the mesh grid for interpolation
X, Z = np.meshgrid(
    np.linspace(fields.var1_mesh[0].min(), 
                 fields.var1_mesh[0].max(), 720), 
    np.linspace(fields.var3_mesh[0].min(), 
                 fields.var3_mesh[0].max(), 720)
)
# Interpolate
gasdens_interp = fields.evaluate(
    field='gasdens', time=0, var1=X, var3=Z
)
velocity_interp = fields.evaluate(
    field='gasv', time=0, var1=X, var3=Z
)

# Colormesh of density
plt.pcolormesh(
    X, Z, np.log10(gasdens_interp),
    cmap='Spectral_r'
)
# Streamlines of velocity
plt.streamplot(
    X, Z, 
    velocity_interp[0], velocity_interp[2], 
    linewidth=0.7, density=3.0
)
\end{python}

The gain in this case, in terms of both breadth and clarity of the analysis code, is already significant. The same procedure as in the previous code, carried out with code snippets of your own code or others working with \fargotd\, can easily be as long as a couple of hundred lines of code. Ignoring stylistic issues (eg, the vertical hanging indent used for fitting the horizontal size of the lines of code to the format of this manuscript), in \fargopy\ this procedure takes only 6 lines of code in total and is far more readable and maintainable.

\subsubsection{Validation of 2D interpolation}
\label{app:fargopy_interp2d}

\begin{figure*}
    \centering
    \includegraphics[width=0.95\textwidth]{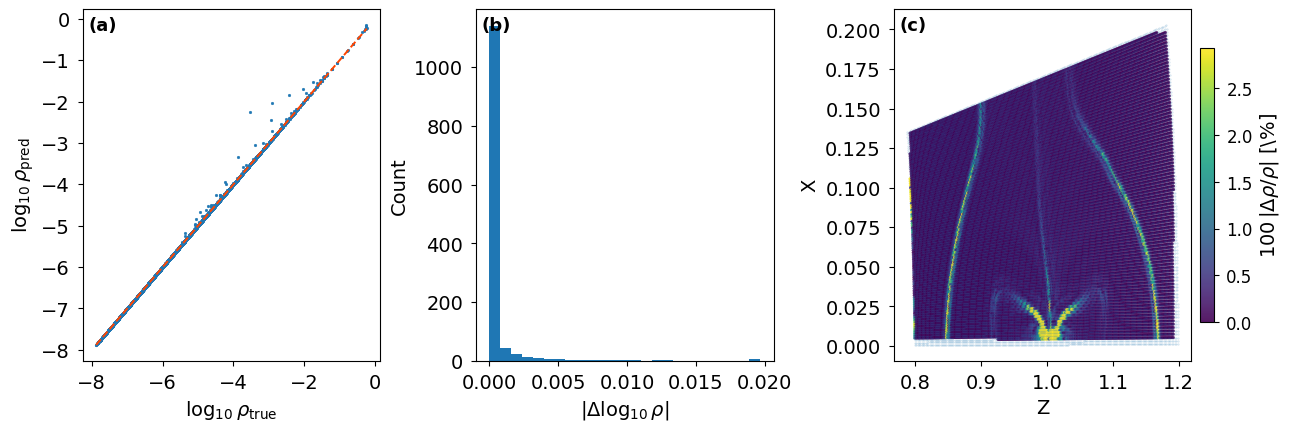}
    \caption{Quantitative diagnostics of the round-trip test applied to the density field. Left: spatial residual maps; right: distribution/summary metrics used to track interpolation accuracy across resolutions and snapshots.}
    \label{fig:metrics2d}
\end{figure*}

To quantify interpolation fidelity, we apply a ``round-trip'' test: interpolate a native 2D field to a regular grid and back to the native mesh, then compute pixel-wise residuals and summary metrics (mean bias, RMS, and robust percentiles). Figure~\ref{fig:metrics2d} summarizes representative diagnostics for a density slice.

\subsubsection{Validation of 3D interpolation}
\begin{figure*}
    \centering
    \includegraphics[width=0.95\linewidth]{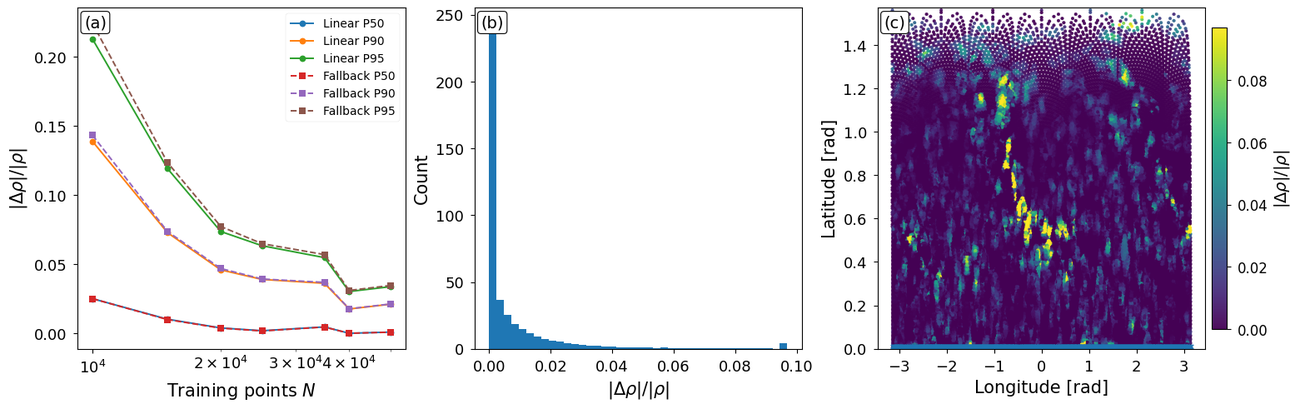}
    \caption{Validation of 3D interpolation of the gas density field on a tessellated spherical surface for a high-resolution simulation.}
    \label{fig:interp3d}
\end{figure*}

We validate the 3D interpolation and surface sampling by comparing interpolated values on the tessellated sphere against the native grid in controlled cases (including high-resolution runs) and by monitoring the convergence of surface-integrated quantities with increasing tessellation refinement. Figure~\ref{fig:interp3d} shows an example validation map for gas density.

\subsection{Surface tessellation}
\label{app:fargopy_tessellation}

Several 3D diagnostics are naturally expressed as integrals over spherical (or nearly spherical) surfaces around an embedded object (e.g., a planet). \webfargopy\ implements a lightweight surface tessellation engine (triangle mesh) that supports:
\begin{itemize}
    \item Sampling of 3D fields on a spherical surface at a prescribed radius (e.g., Hill radius);
    \item Consistent surface elements $\mathrm{d}S$ for integration;
    \item Straightforward visualization of maps on the sphere.
\end{itemize}
\begin{figure*}
    \centering

    \begin{minipage}{0.48\textwidth}
        \centering
        \includegraphics[width=\linewidth]{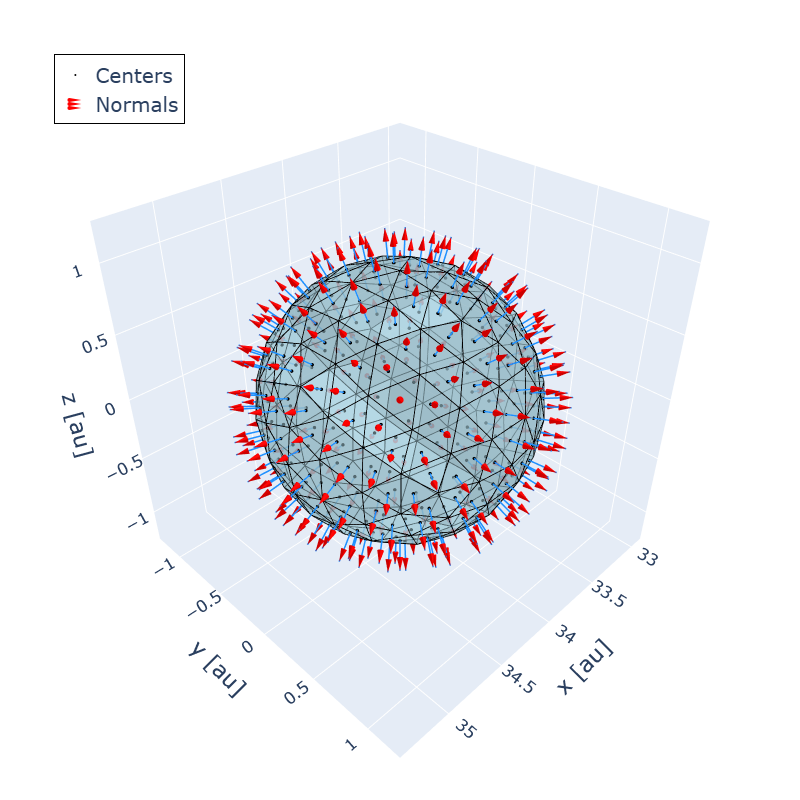}
        \vspace{-0.2cm}
        {\small (a)}
    \end{minipage}
    \hfill
    \begin{minipage}{0.48\textwidth}
        \centering
        \includegraphics[width=\linewidth]{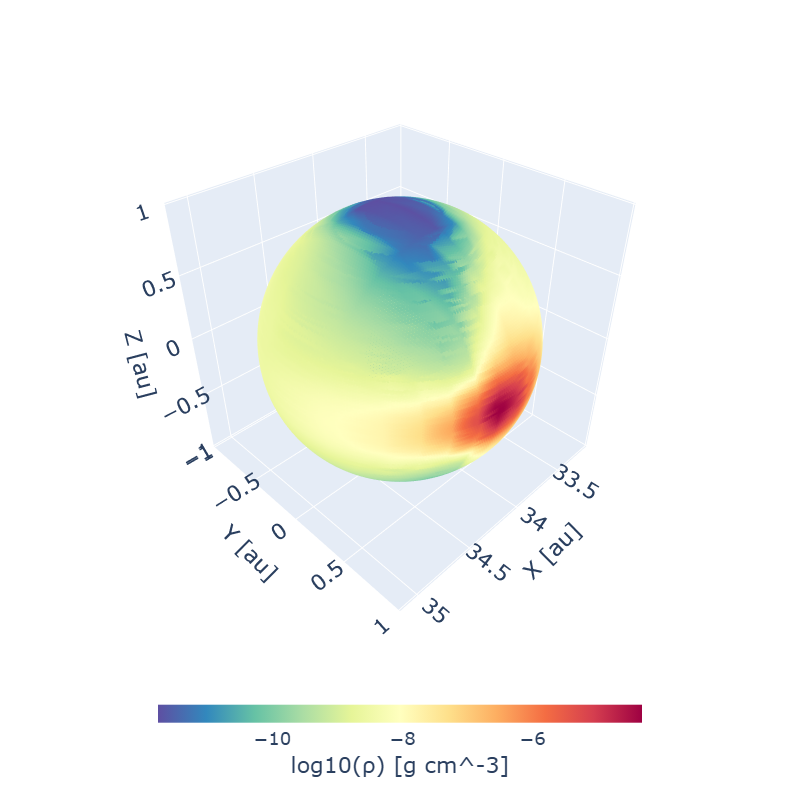}
        \vspace{-0.2cm}
        {\small (b)}
    \end{minipage}

    \caption{
    Panel (a): Construction and use of a tessellated spherical surface to sample 3D fields in the vicinity of an embedded object. The triangular mesh provides both visualization and accurate surface integration via \(dA\). Panel (b): the gas density field interpolated and projected onto the same spherical surface.
    }
    \label{fig:sphere_density_normals}
\end{figure*}

\subsection{Enclosed mass and mass flux}
\label{app:fargopy_mass_integral}

Given a tessellated surface, \webfargopy\ can compute (i) enclosed mass proxies (by combining surface sampling with appropriate geometric factors and/or volume elements, depending on the diagnostic) and (ii) mass flux through the surface using \(\dot{M}=\int \rho\,\mathbf{v}\cdot d\mathbf{A}\). These quantities are used in the main text to characterize accretion and flow morphology around the planet.

\noindent Minimal example: define a spherical surface around the planet and sample fields:
\begin{python}
simpath = fp.Simulation.download_precomputed('p3disoj')
sim = fp.Simulation(output_dir=simpath)

# Load planet properties
planet = sim.load_planets(snapshot=10)[0]

# Define spherical surface at the Hill radius
sphere = fp.Surface(
    type="sphere",
    center=(planet.pos.x, planet.pos.y, planet.pos.z),
    radius=planet.hill_radius,
    subdivisions=6,
    z_cut=0.0
)

# Compute enclosed gas mass
mass = sphere.total_mass(
    sim,
    snapshot=[0,10],
    follow_planet=True
)

# Compute accretion rate
flux = sphere.mass_flux(
    sim=sim, 
    snapshot=[0,10],
    follow_planet=True
) 
t = np.linspace(0,10,11)
plt.plot(t,flux)
\end{python}

Note that it is not necessary to load the data from the density field to perform this calculation: \webfargopy\ automatically reads the required data (the density in this case) and uses it to compute the total mass. The power of \webfargopy\ is demonstrated here not only in organizing the output data from \fargotd\ but also in carrying out truly complex physical calculations with just a few lines of code.

\printcredits

\bibliographystyle{els}
\bibliography{references.bib}

\end{document}